**Predicting the spatial distribution and demographics of commercial swine farms in the United States.**

Felipe E. Sanchez[a;b], Thomas A. Lake[b], Jason A. Galvis[a], Chris Jones[b], Gustavo Machado[a;b] *

[a] Department of Population Health and Pathobiology, College of Veterinary Medicine, North Carolina State University, Raleigh, NC, 27695, USA.

[b] Center for Geospatial Analytics, North Carolina State University, Raleigh, NC, 27607, USA

**\*Correspondence:** Gustavo Machado, gmachad@ncsu.edu

## Abstract

Data on livestock farm locations and demographics are essential for disease monitoring, risk assessment, and developing spatially explicit epidemiological models. In the United States, however, farm-level information is confidential, limiting access to comprehensive national datasets and necessitating the use of predictive tools to estimate farm infrastructure and populations. In this study, we developed a four-stage machine learning framework to detect and characterize commercial swine farms using high-resolution (0.6 m) aerial imagery across two U.S. pig-production regions. First, a deep learning model was trained to identify individual swine barns from background structures using semantic segmentation. Next, we developed a multi-stage filtering pipeline consisting of a Random Forest classifier, a geometric filter, and an OpenStreetMap (OSM)-based filter to refine barn predictions from the semantic segmentation model. Retained barn predictions were subsequently grouped into farms and classified into one of four production types: sow, nursery, finisher, or boar stud, using a Random Forest classifier. Finally, farm-level characteristics and production type labels were used to train a Random Forest regression model to estimate farm-level swine populations. Our semantic segmentation model achieved an F2-score of 92% and a mean Intersection over Union of 76%. An initial total of 194,474 swine barn candidates were identified in the Southeast (North Carolina = 111,135; South Carolina = 37,264; Virginia = 46,075) and 524,962 in the Midwest (Iowa = 168,866; Minnesota = 165,714; Ohio = 190,382). The post-processing Random Forest classifier reduced false positives by 82% in the Southeast and 88% in the Midwest, resulting in 45,580 confirmed barn polygons. These were grouped into 16,976 predicted farms and classified into one of the four production types. Population sizes were then estimated using a Random Forest regression model, with prediction accuracy varying across production types. Across all farms, 87% of predictions for operations with 1,000–2,000 pigs were within 500 pigs of the reference value, with nursery farms showing the highest agreement ($R^2 = 0.82$), followed by finisher farms ($R^2 = 0.77$) and sow farms ($R^2 = 0.56$). Our results revealed substantial gaps in the existing spatial and demographic data on U.S. swine production. While our total pig population estimates exceeded those reported by the USDA, this suggests that the actual number of farms and animals may be underreported in current federal datasets. These findings highlight the potential of integrated machine learning workflows to improve the accuracy and completeness of national-scale swine farm surveillance systems and to serve as a foundation for epidemiological models that rely on such data.

**1. Introduction**

Epidemiological models that rely on incomplete population data limit our ability to quantify disease transmission dynamics and develop effective disease control and eradication strategies

(USDA, 2024; SHIC, 2023; Robinson et al., 2022; Montefiore et al., 2022; Gilbertson et al., 2022; Sellman et al., 2020; Maroney et al., 2020; Handan-Nader and Ho, 2019; Perez et al., 2019, 2015). In livestock systems, information on the location and demography (i.e., herd population and production type) of farms is crucial, as transmission dynamics are driven by farm-to-farm proximity and both direct (animal-to-animal) and indirect (e.g., vehicles moving animals, feed, equipment, or personnel) contacts among the different farms in the production cycle (Cardenas et al., 2023; Sanchez et al., 2023; Galvis et al., 2022, 2021; Niederwerder, 2021; Jara et al., 2021; Büttner and Krieter, 2020; Sanhueza et al., 2020; Dee et al., 2020; Thakur et al., 2016; Pitkin et al., 2009). Furthermore, transmission routes and force of infection differ across regions with varying densities of livestock operations, size of operations, and types of farms (Cardenas et al., 2024; Sykes et al., 2023; Galvis et al., 2022, 2021; Machado et al., 2020). However, farm and demographic data are often unavailable or obfuscated due to confidentiality restrictions and data use agreements (Galvis et al., 2022, 2021; Paploski et al., 2021; Makau, Paploski et al., 2021). Without farm location and demographic data, assumptions about the distribution and size of livestock operations increase uncertainty in the ability to predict transmission rates and forecast when and where the disease will spread, ultimately affecting the accuracy of epidemic models and decreasing the utility for decision-makers and producers (Cardenas et al., 2023; Sanchez et al., 2023; Sykes et al., 2023; Gilbertson et al., 2022; Burdett et al., 2015).

      Recent efforts to estimate livestock population distributions have relied on probabilistic and microsimulation models that use aggregate or imputed data (Burdett et al., 2015; Bruhn et al., 2012; Emelyanova et al., 2009; Neumann et al., 2009). While these models are valuable for identifying broad-scale patterns of livestock distributions, they do not provide the precise

location data needed for epidemiological modeling. For example, the Farm Location and Agricultural Production Simulator (FLAPS) estimates swine farm distributions across the contiguous U.S. by imputing missing data from the Census of Agriculture (CoA) (USDA, 2024), predicting the geographical distribution of individual farms, and simulating swine farm populations (Burdett et al., 2015). Although such models effectively predict demographic trends at broader administrative levels, they do not pinpoint the locations of individual farms. Instead, they simulate where farms may occur, often resulting in discrepancies between predicted and actual farm sites (Burdett et al., 2015). Consequently, while these models are useful for assessing livestock distribution at large scales, they may not accurately capture farm-level spatial patterns, potentially leading to over- or underestimation of epidemic risk.

Remotely sensed imagery combined with machine learning algorithms presents a promising alternative for obtaining farm location and demographic data (Gilbert et al., 2022; Chugg et al., 2021; Handan-Nader et al., 2021; Patyk et al., 2020; Maroney et al., 2020; Handan-Nader and Ho, 2019). Remote sensing provides high-resolution, continuous imagery over large spatial extents, enabling the detection and analysis of farms at scales impractical with traditional field-based methods (e.g., surveys). Machine learning, in turn, enables the analysis of these vast datasets by efficiently identifying patterns and extracting meaningful information from complex imagery. For instance, one study utilized aerial imagery and an object-based image feature extraction technique to detect poultry operations across seven southeastern U.S states (Maroney et al., 2020). While this approach successfully identified poultry operations, it struggled to distinguish them from visually similar buildings, resulting in occasional misclassifications. A subsequent study refined this method by integrating FLAPS data with the machine learning detection model, creating a hybrid approach that incorporated additional spatial and contextual features to improve farm identification (Patyk et al., 2020). Nonetheless, this approach still relies on CoA data, which depends on voluntary participation (Patyk et al., 2020; USDA, 2024).

Deep learning has emerged as a powerful approach for identifying and classifying objects in remotely sensed imagery, outperforming traditional machine learning algorithms such as Random Forest and Support Vector Machines in agricultural and livestock applications (Robinson et al., 2022; Mei et al., 2022; Shea et al., 2022; Chugg et al., 2021; Handan-Nader et al., 2021; Lee et al., 2021; Patyk et al., 2020; Knopp et al., 2020; Handan-Nader and Ho, 2019). Unlike traditional machine learning models, convolutional neural networks (CNNs) are a type of deep learning algorithm that can learn directly from image data and extract spatial features such as edges, textures, and structural patterns. This capability makes CNNs particularly well-suited for identifying barns from remotely sensed imagery, where distinguishing livestock facilities from other structures requires recognizing characteristic layouts and contextual features. Previous studies have demonstrated the effectiveness of CNNs in identifying livestock facilities from aerial imagery. A CNN-based model was trained to detect and classify poultry and swine farms in North Carolina, showing that deep learning models can successfully differentiate these facilities from other structures (Handan-Nader and Ho, 2019). Building on this approach, another study applied semantic segmentation—a deep learning technique using CNNs to classify individual pixels within an image—along with a filtering process to map poultry farm locations across the U.S. (Robinson et al., 2022). Their work produced the first publicly available national dataset of poultry barn locations, highlighting the potential for CNN-based approaches in large-scale livestock facility mapping. Despite these advancements, scaling deep learning models for national-level livestock mapping remains challenging. Training CNNs requires extensive annotated datasets, which are costly and time-consuming to produce, and require substantial computational resources. Overcoming these limitations is essential to fully harness deep learning for accurate, high-resolution mapping of livestock facilities.

In this study, we developed a four-stage machine learning framework to predict the locations of commercial swine farms, production types, and population sizes across the six pig-producing states in the U.S., including North Carolina, South Carolina, Virginia, Iowa, Minnesota, and Ohio. Using high-resolution (0.6 meters) aerial imagery from the National Agriculture Imagery Program (NAIP) and data from the Rapid Access Biosecurity application (RABapp™) repository (RABapp, 2024), we first implemented a deep learning model to identify individual barns through semantic segmentation. To refine these predictions, we developed a multi-stage filtering pipeline consisting of a Random Forest classifier, a geometric filter, and an OpenStreetMap (OSM)-based filter that incorporated geometric, spatial, and environmental features to distinguish swine barns from non-barn structures. Following this filtering step, a second Random Forest classifier was used to categorize predicted barns into four swine production types: sow farms, nursery farms, boar studs, and finisher farms. Finally, a Random Forest regression model was applied to estimate farm-level swine populations by integrating barn-level characteristics with available farm metadata (e.g., production type, pig capacity).

## 2. Materials and Methods

### 2.1. Swine production distribution data

The U.S. swine industry comprises approximately 131,471 farms housing 75.8 million pigs (USDA NASS, 2024; USDA, 2024; Passafaro et al., 2020; Reimer, 2006). Based on the 2022 hog inventory reported by the U.S. Department of Agriculture's National Agricultural Statistics Service (NASS), six states including, North Carolina (8.2 million), South Carolina (0.2 million), Virginia (0.3 million), Iowa (24.6 million), Minnesota (9.5 million), and Ohio (2.7 million), collectively accounted for approximately 45.5 million pigs, representing about 60% of the national total (USDA, 2024; SHIC, 2023; USDA NASS, 2024; USDA, 2019). We considered these six major pig-producing states in our study, which span both the Midwestern and Southeastern U.S., to capture diverse production systems and geographic contexts. We categorized swine farms into four production types: sow, nursery, finisher, and boar stud, based on company-provided production type data (Supplementary Material Section S1 and Supplementary Material Table S1).

## 2.2. Imagery datasets and pre-processing

NAIP captures aerial imagery of the U.S. during the agricultural growing season by staggering statewide collections on a three-year cycle (USDA, 2022). For this study, we used 2022 - 2023 NAIP imagery for North Carolina, South Carolina, Virginia, Iowa, Minnesota, and Ohio. Imagery was downloaded from the USDA's Geospatial Data Gateway and consisted of county-level mosaics in either three-band (natural color: red, green, blue) or four-band (natural color and near-infrared) format, with a spatial resolution of 0.6 meters (Figure 1-a) (USDA, 2022). We then pre-processed the county-level aerial imagery by extracting non-overlapping 512 × 512-meter image tiles for each county (Figure 1-b).

*2.3. Rapid Access Biosecurity application (RABapp™) repository*

The RABapp™ database includes detailed biosecurity information on farm features used to create standardized biosecurity maps of swine production premises in compliance with Secure Pork Supply (SPS) plan guidelines (Supplementary Material Figure S1 and Figure 1-a) (Secure Pork Supply, 2025; Fleming et al., 2025). Among map biosecurity features, the line of separation (LOS) delineates the boundary surrounding buildings that house animals and areas where employees and equipment undergo cleaning and disinfection procedures (Supplementary Material Figure S1) (Fleming et al., 2025). For this study, we used LOS from 5,160 commercial swine farms in the RABapp™ database spanning 22 states to delineate individual barn polygons (Supplementary Section S2 and Supplementary Material Figure S2) (Fleming et al., 2025). Although the final predictions focused on six major swine-producing states: North Carolina, South Carolina, Virginia, Iowa, Ohio, and Minnesota, we included data from Alabama, Arkansas, Colorado, Delaware, Georgia, Illinois, Indiana, Kansas, Michigan, Mississippi, Missouri, Nebraska, Oklahoma, Pennsylvania, Texas, and Wyoming during model training to increase the diversity and volume of examples across a wide range of environments (Adegun et al., 2023; Yuan et al., 2021; Shorten and Khoshgoftaar, 2019; Garcia-Garcia et al., 2018).

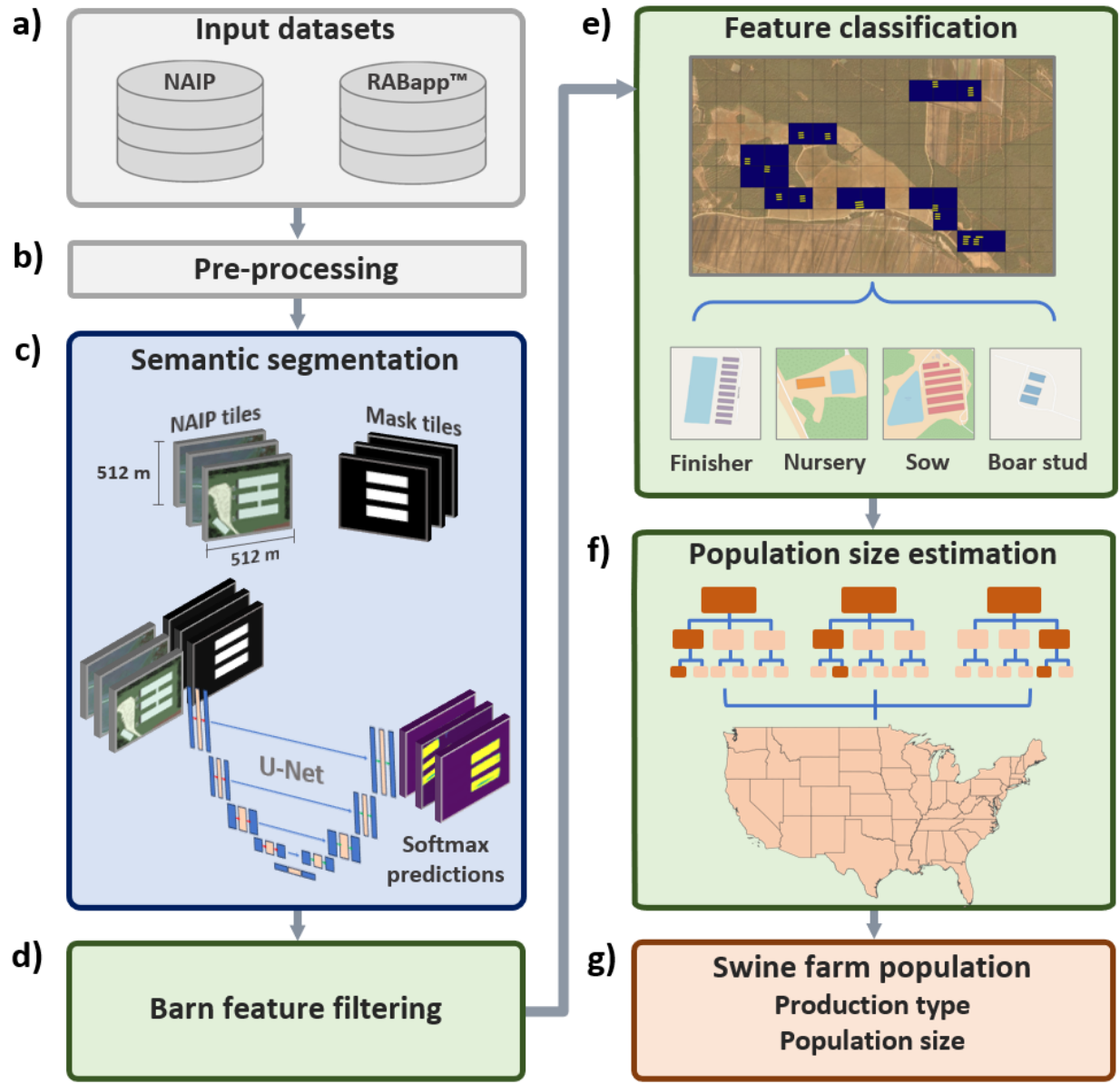

**Figure 1. Conceptual diagram of workflow.** Directional workflow used in this study, where the input datasets (a), consisting of NAIP imagery and RABapp™ barn polygons, were pre-processed (b), by clipping NAIP imagery into 512 m² tiles and generating corresponding masks from RABapp polygons for input into a semantic segmentation model (c). Predictions obtained from the semantic segmentation model underwent a feature filtering process (d) to eliminate false positives from our predicted datasets before applying our Random Forest classification model (e). Once swine barns were classified into production types (finisher,

nursery, sow, and boar stud), the population size for each barn was estimated using a Random Forest regression model (f). Finally, we stitched together all the model predictions to produce a final map of swine barns across North Carolina, South Carolina, Virginia, Iowa, Ohio, and Minnesota (g). The blue box represents our deep learning model, and the dark green boxes represent the Random Forest models. The orange box represents our final product

*2.4. Binary semantic segmentation of barns*

We developed a semantic segmentation model based on the U-Net architecture to classify pixels from NAIP imagery into "barn" and "background" classes (Ronneberger et al., 2015; Yuan et al., 2021) (Figure 1-c). The U-Net model features two distinct paths: a contracting path and an expansive path (Ronneberger et al., 2015). The contracting path extracts contextual information (e.g., edges, shapes, textures) from the image through convolution, activation, and pooling layers. The expansive path then decodes this information using layers that project the features back to the original image dimensions (Ronneberger et al., 2015; Yuan et al., 2021). We integrated a Residual Network (ResNet) backbone into the U-Net contracting path, allowing for effective information transfer between layers, which enhances feature extraction and representation, and improves overall model performance (He et al., 2015). This configuration has been shown to outperform other model architectures used for building footprint extraction tasks (Aghayari et al., 2023; Sariturk and Seker, 2023; Stiller et al., 2023). Specifically, we employed a ResNet-50 encoder pre-trained on the ImageNet dataset, which enables the model to utilize prior knowledge of a wide range of features from a diverse collection of images (Sariturk and Seker, 2023; Krizhevsky et al., 2017; He et al., 2015). This pre-trained encoder serves as a baseline for the model to recognize commercial swine barns. The final layer of our U-Net model uses the Softmax function to normalize the outputs on a scale of [0,1], generating predicted probabilities for each class at the pixel level (Gao and Pavel, 2017).

The U-Net model was trained and evaluated using 9,486 image-mask tile pairs containing swine farms from the RABapp™ database (Table 1 and Figure 4). To aid in model generalizability, we allocated 70% of the tiles for model training (n = 6,640), 10% for validation (n = 949), and 20% for testing (n = 1,897) from swine farms spanning 22 states (Supplementary Material Figure S3) in the contiguous U.S. In addition to the 6,640 image-mask tile pairs containing swine farms, we included 2,013 image-mask tile pairs with no swine barns (true negatives), bringing the total of training image-mask tile pairs to 8,653. We applied image augmentation techniques to further support model generalization, including discrete rotational transformations (0°, 90°, 180°, and 270°), and hue, saturation, and value transformations (Farahnakian et al., 2023; Goodfellow et al., 2016). These augmentations exposed the model to a broader range of scenarios during training, enhancing its adaptability to unseen data. Despite the study's focus on regions of high pig production, the swine barn class is still considered a rare feature, constituting only 1.15% (196,635,089 pixels) of the 17,048,429,968 pixels in our dataset. To address class imbalance, we calculated class-specific weights based on pixel frequencies, assigning a weight 15 times greater to barn pixels compared to background pixels (15.35 for barn pixels and 0.51 for background pixels). The class-specific weights were incorporated into a custom weighted binary cross-entropy loss function that assigns the true class weight to each pixel's error. The model was trained for 100 iterations of the training dataset (epochs) using the Adaptive Moment Estimation (Adam) optimizer, and an initial learning rate of 0.0001 (Farahnakian et al., 2023; Goodfellow et al., 2016; Kingma and Ba, 2014). To address overfitting, we incorporated a learning rate decay that reduced the learning rate by a factor of ten when training loss plateaued, and an early stopping callback that halted the training process after fifteen consecutive epochs of stagnating performance.

We evaluated the semantic segmentation model's performance using metrics calculated on the withheld (20%, n = 1,987 image-mask pairs) testing dataset. Performance metrics included overall accuracy, precision, recall, specificity, mean intersection over union (IoU), and the F2-score. Overall accuracy was measured as the proportion of correctly predicted pixels (both barns and background) among all predictions. Precision quantified the proportion of pixels correctly predicted as barns among all pixels classified as barns, whereas recall assessed the model's ability to detect actual barn pixels. Specificity evaluated the model's effectiveness in identifying background pixels (true negatives). Mean IoU measures the overlap between predicted and reference barn pixels relative to their combined area. Finally, the F2-score emphasizes recall more than the conventional F1-score, making it particularly suitable for scenarios where missing a potential object of interest (false negative) carries a higher cost than predicting a false positive. The formulas used to calculate these metrics are provided in Supplementary Materials Section S3. While 512 × 512-meter tiles were used for model training and testing, final state-level predictions were performed using larger 2048 × 2048-meter tiles. This approach follows the methodology proposed by Huang et al. (2018), which demonstrated that using larger input tiles during inference mitigates artifacts at tile edges and improves segmentation accuracy by reducing translational variance and edge-related errors.

## 2.5. Random Forest barn feature filtering

Swine barn feature filtering was carried out by a Random Forest classification model that distinguishes swine barns from non-swine barn features based on geometric, spatial, and environmental characteristics (Figure 1-d). First, we generated binary predictions at the pixel level by applying a threshold value of 0.7 to the semantic segmentation model's probabilistic output (softmax) predictions (Garcia-Garcia et al., 2018) (Figure 1-c). This threshold was fine-tuned using the 949 image-mask tile pairs in the validation dataset by identifying the value that maximized the mean IoU across a range of thresholds tested in 0.1 increments from 0.1 to 1.0 (Supplementary Material Figure S4). Contiguous pixels predicted as the positive barn class were grouped into polygons using an 8-connected pixel approach, ensuring that only connected clusters of barn-class pixels were treated as individual polygons (Supplementary Material Figure S5). This approach provided a computationally efficient method suitable for capturing the predominantly rectangular shape of swine barns (Chaudhuri and Samal, 2007).

Predicted polygons were manually reviewed, and we selected 22,510 representative false positives for the Southeast (NC = 4,916; SC = 9,981; VA = 7,622) and 12,314 for the Midwest (IA = 3,108; MN = 4,736; OH = 4,470). These sets correspond to approximately 11% of all predictions in the Southeast (194,474 total) and 2.5% in the Midwest (492,584 total) and were sampled across multiple counties in every state to capture a diverse range of false positives (e.g., parking lots, houses, malls, and warehouses) and land-use contexts. These false positives were combined with the RABapp™ barns data from each of the six states to train the Random Forest classification model to filter out non-barn features (Table 1). For each polygon in the combined dataset, we calculated the area (m²) representing the total enclosed surface of the polygon, and the aspect ratio as the width divided by the length of the polygon's bounding box, which provided a geometric descriptor that captures the rectangular proportions of each polygon (barn) (Supplementary Material Figure S5). Next, we downloaded TIGER/Line shapefiles for primary and secondary roads for each state (U.S. Census Bureau., 2023) and calculated the

Euclidean distance from each polygon to the nearest primary or secondary road. Landscape composition was incorporated into the dataset using the 2023 National Land Cover Database (NLCD) land cover raster product (United States Geological Survey, 2023). We applied buffers of 500 m, 1,000 m, and 5,000 m around each feature in our combined dataset to extract land cover class proportions at multiple scales. The proportions within each buffer were assigned as separate columns, providing a multi-scale representation of the surrounding environment. To account for regional variation in land cover composition (Supplementary Material Figure S7), we trained separate Random Forest models for the Southeastern states (NC, SC, VA) and the Midwestern states (IA, MN, OH). We evaluated each regional model using a 5-fold spatial cross-validation strategy. Within each region, the study area was partitioned into 25 x 25 km spatial blocks, and blocks containing barns and false positives were randomly allocated to one of five folds (Supplementary Material Figures S8 and S9). For each iteration, four folds were used for training and one for testing, ensuring spatial independence between the training and testing sets. A summary of the sample distribution across folds within each region, including the number of barns and false positives used in each split, is provided in Supplementary Material Tables S2 and S3, and Supplementary Material Figures S8 and S9.

For each buffer distance used to extract land cover class proportions (500 m, 1,000 m, and 5,000 m), we applied a Grid Search within each spatial fold to identify the optimal set of hyperparameters. The search systematically evaluated combinations of parameters, including the number of trees (100, 200, 300, 500), maximum tree depth (None, 10, 20, 30), minimum samples required to split an internal node (2, 5, 10), minimum samples required at a leaf node (1, 2, 4), and the number of features considered at each split ('sqrt', '$\log_2$') (Yang and Shami, 2020; Pedregosa et al., 2011). The best hyperparameter combination for each fold and buffer distance was selected based on the F1-score and used to train that fold independently. In total, 15 models were trained and evaluated per region (5-fold cross-validation across 3 buffer distances). Model performance was evaluated as the mean of test metrics obtained across all folds (Supplementary Material Tables S4 and S5). For each fold, the model assigned a probability score (0–1) to every polygon in the test set, representing the likelihood that it was a true swine barn. Final classifications were determined using a voting-based approach, where polygons with predicted probabilities ≥0.5 in at least three out of five folds were retained as barns; otherwise, they were classified as false positives (Brown, 2017). Feature importance scores, calculated using the Gini index (mean decrease impurity), were averaged across folds and analyzed by buffer distance to assess how variable influence changed with spatial scale across the six states and two regions (Pedregosa et al., 2011) (Supplementary Material Figure S10 and S11). Once the final classifications were made using the Random Forest model, we applied two geometric filters. First, we removed overlapping polygons (duplicates) introduced by tiling and mosaicking, leaving only a single representative polygon per overlapping cluster. Second, we filtered by size using empirical bounds from the RABapp™ dataset. Polygons with areas below the 10th percentile ($q10 = 500$ m²) or above the 90th percentile ($q90 = 5,000$ m²) were excluded. Lastly, we filtered our predictions using OpenStreetMap© (OSM) building and road data for each state extracted using the OSMnx Python package (Boeing, 2025). Polygons intersecting building features tagged with clearly non-barn uses (e.g., schools, churches, warehouses, industrial) were excluded, whereas those intersecting generic/agricultural tags (e.g., yes, farm_auxiliary, farm, barn) or swine-specific tags (sty) were retained. For roads, polygons were removed only when the intersection involved major highways (motorway, trunk, primary, including link variants). A complete list of tags and counts of intersecting features is provided in the Supplementary Materials, Tables S8-S11.

## 2.6. Random Forest farm classification model

Using the RABapp™ dataset of barn polygons, we implemented a Random Forest classification model to categorize swine farms into four production types: sow, nursery, finisher, and boar stud. Barn-level data from the RABapp™ dataset were consolidated back to the farm level (n = 5,160 farms) by using the original farm identifier provided by the company. Once aggregated back to the farm level, we calculated the following predictor variables: mean and standard deviation of barn area, aspect ratio, width, length, and the number of barns per farm. The aggregated dataset was then partitioned into training (70%, n = 3,612) and testing (30%, n = 1,548) subsets. A 5-fold cross-validation Grid Search was used to optimize the Random Forest classifier (Yang and Shami, 2020; Pedregosa et al., 2011). Performance metrics included accuracy, precision, recall, and the F1-score to assess the model's effectiveness in classifying swine farms by production type.

Since the outputs from steps c through d (Section 2.3 through Section 2.5) consisted of individual barn polygons, we aggregated these predictions to represent swine farms. Predicted barn polygons located within 500 meters of one another were grouped to form individual farms. We then applied the trained Random Forest on the newly aggregated predicted farms generated from the filtered, predicted barn polygons (Section 2.5) from each state.

Table 1. Summary of datasets and sample sizes used across model development steps

| Step | Purpose | Data source | Extent | Sample type | Training | Validation | Test |
|---|---|---|---|---|---|---|---|
| Fig. 1 step c: Semantic segmentation | Barn detection | NAIP imagery + RABapp™ | 22 states | Image and mask tile pairs (512 m²) | 8,653 image and mask pairs | 949 image and mask pairs | 1,897 image and mask pairs |
| Fig. 1 step d: Feature filtering (Southeast) | Remove false positives (barn level) | RABapp™ barns + manually identified false positives | NC, SC, VA | Individual polygons | Supplementary Material Table S2 | | |
| Fig. 1 step d: Feature filtering (Midwest) | Remove false positives (barn level) | RABapp™ barns + manually identified false positives | IA, MN, OH | Individual polygons | Supplementary Material Table S3 | | |
| Fig 1. step e: Production type classification | Classify farms into production types | RABapp™ derived farms | NC, SC, VA, IA, MN, OH | Farm level records | 3,612 farms | — | 1,548 farms |
| Fig. 1 step f: Population size estimation | Predict farm-level population | RABapp™ farms + classified predicted farms | NC, SC, VA, IA, MN, OH | Farm level records | 4,119 farms | — | 1,030 farms |

**Note:** A dash (—) indicates that no samples were used for that component of the workflow.

*2.7. Farm-level swine population predictions*

Population size estimates were generated using a Random Forest regression for each predicted swine farm generated from steps c through e (Figure 1 and Table 1). Given that the RABapp™ dataset contains population data at the farm level, we utilized the previously aggregated dataset (Section 2.6) of 5,160 farms, derived by consolidating barn-level data, for model training and testing. Farms without reported population data (n = 11) were excluded from the analysis, resulting in a final sample of 4,119 farms for training and 1,030 farms for testing (Table 1). The model also incorporated previously calculated metadata in Section 2.6, including the mean and standard deviation of barn area, length, width, aspect ratio, the number of barns per farm, and production type. Hyperparameter optimization was performed using Grid Search with 5-fold cross-validation, evaluating combinations of hyperparameters, including the number of trees, maximum tree depth, minimum samples required to split an internal node, and minimum samples required to reach a leaf node (Yang and Shami, 2020; Pedregosa et al., 2011). The Grid Search was used to evaluate the coefficient of determination ($R^2$) and root mean squared error (RMSE), with model selection based on maximizing $R^2$. The final Random Forest regression model, configured with optimal hyperparameters, was evaluated on our withheld testing dataset. The trained Random Forest regression was then applied to the predicted farms obtained from the steps c through e (Section 2.3 through Section 2.6) in our model workflow (Figure 1c-e). Farm-level population estimates were generated for each region and benchmarked against USDA-reported data (USDA, 2024) at the state level to assess the alignment and reliability of the model's predictions.

## 3. Results

### 3.1. RABapp™ barn descriptive analysis

A total of 19,636 barns were extracted from 5,160 commercial swine farms spanning 22 states in the RABapp™ database (Supplementary Material Figure S2 and S3) (Fleming et al., 2025; RABapp, 2024). The number of barns per farm varied by production type, with sow farms having the highest median number of barns per farm (5; IQR: 3–8), followed by finisher farms (2; IQR: 2–4), nursery farms (2; IQR: 1–4), and boar stud farms (2; IQR: 1–3). Farm layout differed across production types, with boar stud farms exhibiting the highest proportion of single-barn farms (46%), followed by nursery (33%), finisher (20%), and sow farms (3%). Barn size and shape also varied by production type, with sow barns being the largest (median area: 1,255 m²; IQR: 742–2,100 m²), followed by finisher (828 m²; IQR: 734–1,242 m²), nursery (615 m²; IQR: 482–922 m²), and boar stud (521 m²; IQR: 349–803 m²). Finisher barns were the most elongated in shape with an aspect ratio of 2.44 (IQR 1.73–3.62), followed by sow (2.12; 1.54–3.18), boar stud (1.80; 1.28–2.88), and nursery (1.56; 0.87–2.15).

Sow farms contained the most spatially dispersed barns, with a median barn-to-barn centroid distance, named hereafter as intra-barm distance[1], of 67 meters (IQR: 46–89 m; max: 742 m), followed by finisher farms (49 m; IQR: 36–69 m; max: 1,276 m), boar stud (47 m; 39–71 m; max 3,571 m), and nursery farms (42.0 m; 29.6–82.2 m; max 1,302.0 m). Overall, 99% of all multi-barn farms had median intra-barn distances under 500 m. In the Southeast, Virginia exhibited the highest median intra-barn distance of 70.3 m (IQR:26.1–82.8 m; max 762.1 m), followed by 68.9 m in South Carolina (37.8–83.2 m; max 180.0 m), and 49.8 m in North

---

[1] Intra-barn distances were calculated as the pairwise Euclidean distances between all barn centroids within each farm.

Carolina (IQR 31.3–79.7 m; max 1,302.0 m). In the Midwest, median intra-farm distances were 46.4 m in Iowa (38.2–55.1 m; max 1,276.9 m), 45.2 m in Minnesota (28.1–65.0 m; max 1,039.8 m), and 37.2 m in Ohio (IQR: 29.3–70.9 m; max: 128.3 m).

*3.2. Semantic segmentation of barns*

Our semantic segmentation model achieved an F2-score of 92% and a mean IoU of 85% during model validation (Supplementary Material Figure S6). When applied to our test dataset, the U-Net model achieved an F2-score of 92% and a mean IoU of 76% for identifying swine barns. The model generated a total of 194,474 predicted barn polygons across the Southeastern states (NC = 111,135; SC = 37, 264, VA= 46, 075 ), and 524,962 predicted barn polygons across the Midwestern (IA = 168,866 , MN = 165,714, OH = 190,382) states (Table 2, Supplementary Material Table S6 and S7). Model-predicted barn polygons were, on average, 23% larger than their corresponding RABapp™ labels in the Southeastern states and 24% larger in the Midwestern states (Figure 3a and 3b). This suggests the model overestimated barn boundaries, particularly along edges where transitions between structures and background were less distinct (Figure 2a, 2b).

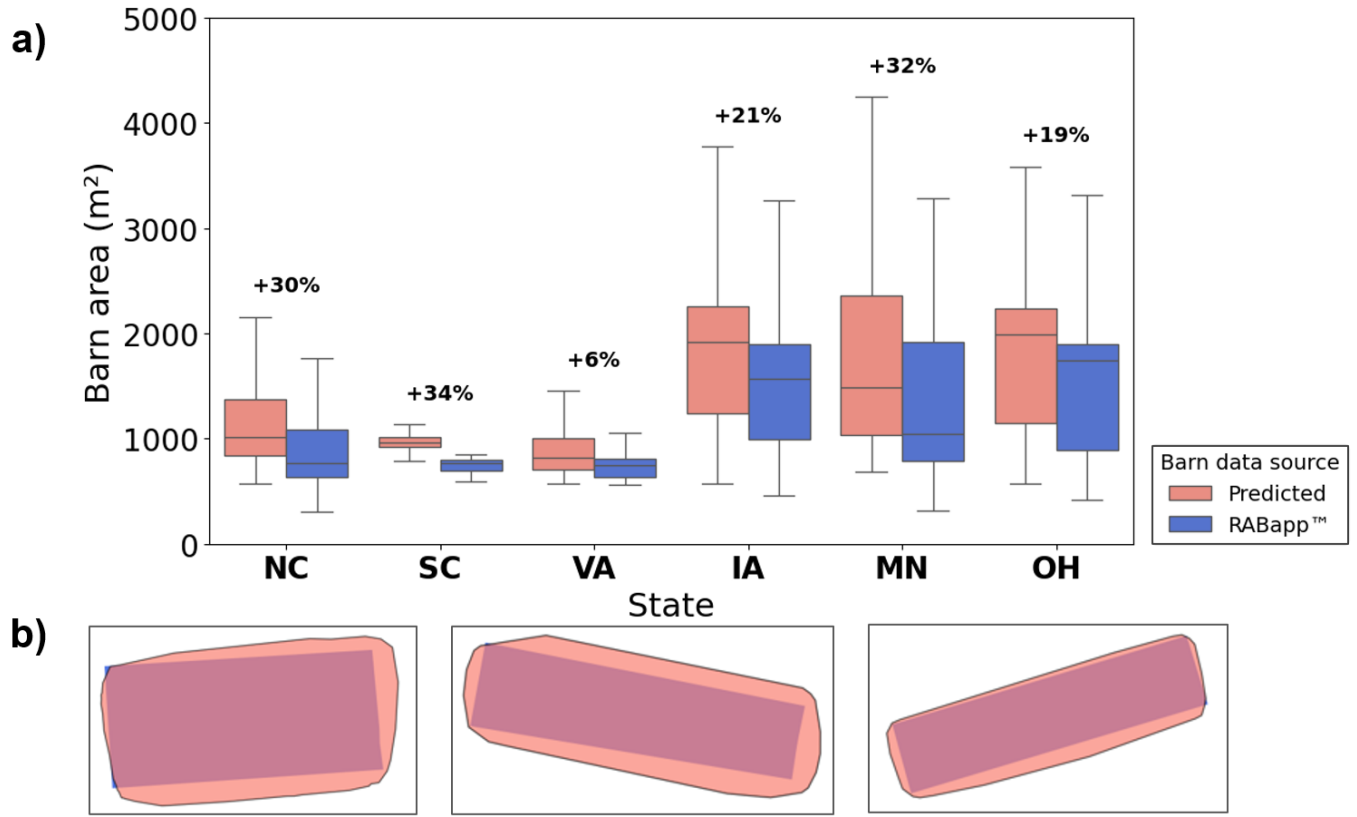

**Figure 2. Comparison of barn area distributions by state for RABapp™ and predicted barns. a)** RABapp™ barns (blue) and model-predicted barn polygons that intersect with RABapp™ barns (red). **b)** Example of model-predicted polygons overlaid on RABapp™ barns.

*3.3. Barn feature filtering*

The Random Forest classification model trained on Southeastern states achieved a mean accuracy of 98% and an average F1-score of 99% across the five spatial folds (Supplementary Material Table S4). The Midwestern Random Forest model achieved an average accuracy of 96% and an average F1-score of 95% across its corresponding five test spatial folds (Supplementary Material Table S5). Across the five spatial folds, the best performing configuration was associated with the 1km buffer in the Southeast, yielding a mean accuracy of 98% and a mean F1-score of 99% (Supplementary Material Table S4), and the 5km buffer in the Midwest with an 96% accuracy and a 95% F1-score (Supplementary Material Table S5). Across all buffer distances, the most influential features contributing to model performance were related to the proportion of pixels designated as "Developed" ("Developed open space", "Developed low intensity", "Developed medium intensity", and "Developed high intensity") and "Cultivated crops" (Supplementary Material Figure S10 and S11).

Applying the 1km Random Forest classifier to all predicted polygons from Section 3.2 reduced the number of predicted barn polygons by 82% (from 194,474 to 34,728) in the Southeastern states (Table 2 and Supplementary Material Table S6). In the Midwestern states, we applied the 5km Random Forest classifier, which reduced the number of predicted polygons by 88% (from 524,926 to 62,669) (Table 2 and Supplementary Material Table S7). Further, we applied the geometric filter to the 34,728 polygons retained for the Southeastern region and removed an additional 17,828 polygons, resulting in 16,900 predicted barns across the Southeast. In the Midwest, the geometric filter was applied to the 62,669 polygons, removing 33,245, leaving 29,424 across Iowa, Minnesota, and Ohio (Table 2). Finally, we filtered both regional datasets using OSM building and road tags from each state. This filtering step removed an additional 243 polygons in the Southeast, bringing the final count of predicted barn polygons to 16,657 for the region. The OSM filter in the Midwest removed an additional 501 polygons, yielding a final total of 28,923 predicted barns across the Midwestern states (Table 2).

**Table 2.** Number of predicted barn polygons from the semantic segmentation model (Section 3.2), and the number of retained polygons after application of filtering steps, including the Random Forest, geometric, and OSM-based filters, for each state in the Southeastern (NC, SC, VA) and Midwestern (IA, MN, OH) regions.

| State | Predicted Polygons | Random Forest filter | Geometric filter | OSM filter |
|---|---|---|---|---|
| NC | 111,135 | 28,648 | 14,106 | 13,894 |
| SC | 37,264 | 4,138 | 1,770 | 1,746 |
| VA | 46,075 | 1,942 | 1,024 | 1,017 |
| IA | 168,866 | 30,286 | 15,611 | 15,403 |
| MN | 165,714 | 25,033 | 10,538 | 10,326 |
| OH | 190,382 | 7,350 | 3,275 | 3,194 |

### *3.4. Production type classification of barns*

We trained a Random Forest classification model to categorize the predicted swine operations into four production types: sow, nursery, finisher, and boar stud. The final model achieved 87% accuracy and an F1-score of 89%. Total barn area and mean barn length were the strongest predictors of production type, followed by measures of barn shape (aspect ratio) and barn width (Supplementary Material Figure S12). Performance was highest for finisher farms (F1-score: 0.94), followed by sow (F1-score: 0.80), and nursery farms (F1-score: 0.79). The boar stud class, which represented less than 1% of the training data, was not correctly identified by the model (F1-score: 0).

To approximate swine farm premises, predicted barns obtained from Section 3.3 were aggregated into individual farms using a 500 m buffer. This distance was selected based on analyses of the RABapp™ dataset (Section 3.1), which showed that 99% of all multi-barn farms had median intra-barn distances below 500 m. We then applied the trained Random Forest classifier to this full set of predicted farms for each state. The resulting classification identified 4,530 farms in the Southeastern region as finisher, 181 as sow, 649 as nursery, and 70 as boar stud. In the Midwestern region, the classifier identified 12,447 farms as finisher, 459 as sow, 3,019 as nursery, and 232 as boar stud for a total of 16,157 predicted farms. Compared to the

RABapp™ data, the predicted proportion of finisher farms, expressed as a percentage of total farms within each state, was overestimated in North Carolina (Predicted: 82% vs RABapp™: 62%) and South Carolina (Predicted: 86% vs RABapp™: 85%), while slightly overestimated in Virginia (Predicted: 87% vs RABapp™: 85%) and underestimated in Iowa (Predicted: 78% vs RABapp™: 97%) and Minnesota (Predicted: 75% vs RABapp™: 80%). In Ohio, the proportion of finisher farms was slightly higher in our predictions (Predicted: 79% vs RABapp™: 76%). Sow and nursery farms showed consistent proportions across most states; however, sow farms were underrepresented in our predictions for North Carolina (Predicted: 4% vs RABapp™: 16%) and Minnesota (Predicted: 3% vs RABapp™: 9%), while slightly overrepresented in Iowa (Predicted: 3% vs RABapp™: 0.3%). Boar stud farms were overestimated by 0.6–1.4% in the Southeastern region (NC: 1.3%, SC: 1.4%, VA: 0.7%) and 1.1 - 1.6% in the Midwestern region (IA: 1.4%, MN: 1.6%, OH: 1.1%) (Figure 3).

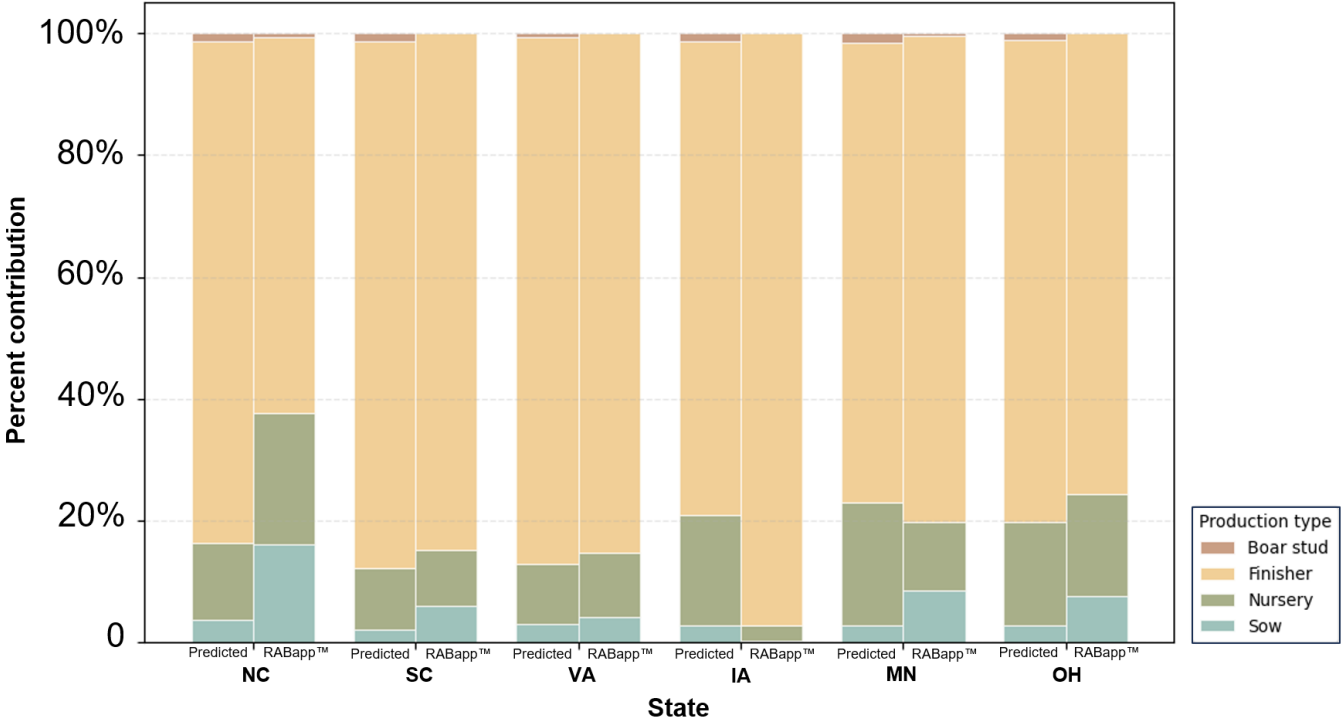

**Figure 3. Distribution of production types by state.** Comparison of swine production type composition between predicted and RABapp™ farms across six states. Bars represent the percent contribution of each production type (sow, nursery, finisher, and boar stud) to the total number of farms identified per source (Predicted vs. RABapp™).

## *3.5. Farm-level population estimates*

The Random Forest regression model was trained to estimate swine farm population sizes based on barn characteristics and production type. The final model achieved an $R^2$ score of 0.77 and a root mean squared error (RMSE) of 1,480 pigs (95% CI: 1,327–1,632). Overall, the model predicted 63% of the farms with up to 500 pigs, 78% of farms with 500 pigs to 1,000 pigs, and 87% of farms with 1,000 to 2,000 pigs. However, performance varied by production type, with finisher, nursery, and sow farms performing well, while boar studs' population prediction underperformed (Figure 4). Specifically, 67% of finisher farms predicted within 500 pigs and 88% within 2,000 pigs ($R^2 = 0.77$), 57% of nursery farms predicted within 500 pigs and 86% were within 2,000 pigs ($R^2 = 0.82$), 55% of sow farms predicted were within 500 pigs and 83% within 2,000 pigs ($R^2 = 0.56$), and 50% of boar stud farms predicted with 500 pigs and 76% within 2,000 pigs (Figure 5). Although the $R^2$ value for boar stud farms was low, the predictions generally followed the observed pattern, which was confirmed by a significant positive Pearson correlation ($r = 0.89$), indicating that the model captured the underlying relationship despite the small number of observations. When aggregated to the state level and compared with USDA Census of Agriculture data (USDA, 2024), predicted values exceeded reported swine inventories in all six states (Table 3). The most significant differences were observed in South Carolina (+28,900%; Predicted: 2.9 million vs. USDA: 0.01 million) and Virginia (+2,600%;

Predicted: 1.6 million vs. USDA: 0.06 million). In contrast, predictions for Iowa (+11%; Predicted: 27.4 million vs. USDA: 24.6 million), Minnesota (+92%; Predicted: 17.7 million vs. USDA: 9.2 million), Ohio (+135%; Predicted: 5.7 million vs. USDA: 2.45 million), and North Carolina (+140%; Predicted: 19.7 million vs. USDA: 8.2 million) were more consistent with national inventory patterns, though they still showed higher swine populations as compared to those reported in the USDA Census of Agriculture (USDA, 2024).

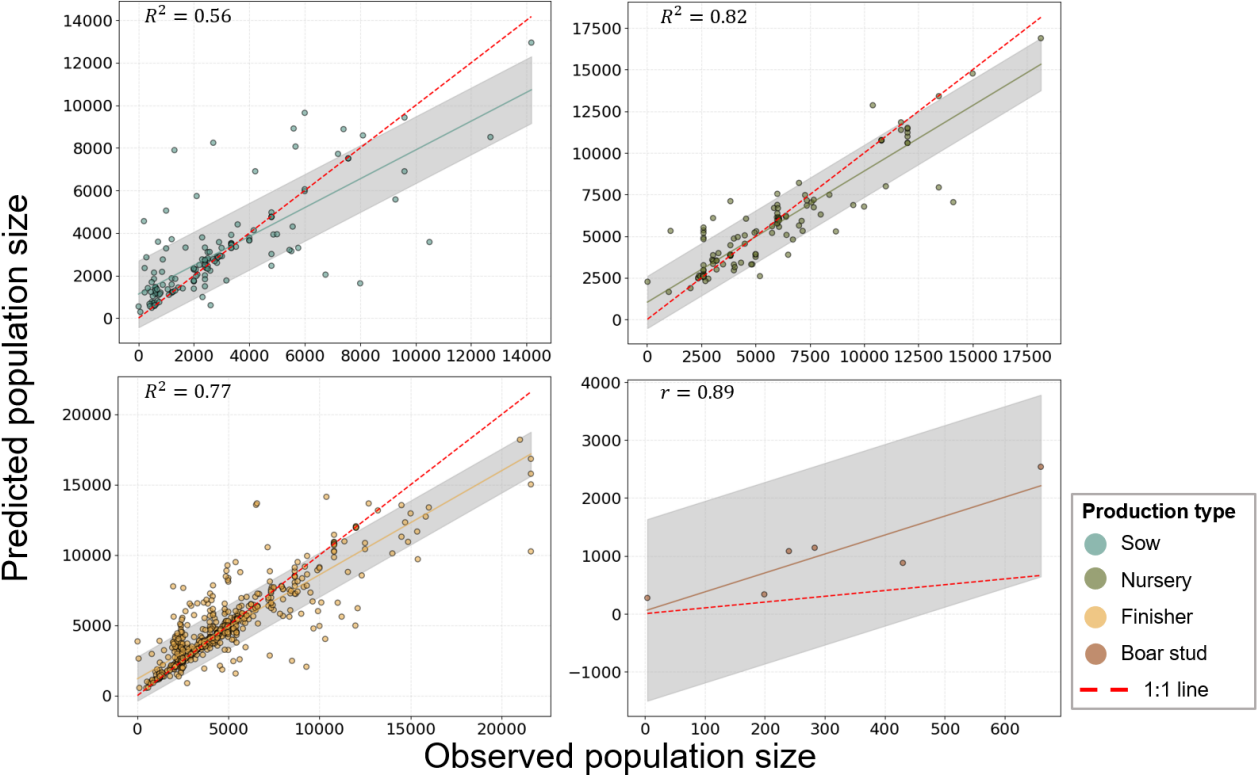

**Figure 4. Scatter plot of actual vs. predicted swine farm population sizes per production type.** Each point represents an individual farm, color-coded by production type (finisher, nursery, sow, and boar stud). The shaded gray area denotes the 95% confidence interval of the root mean squared error (RMSE), indicating the expected range of prediction deviation.

**Table 3.** The number of predicted farms, barns, and population size for three Southeastern states (NC, SC, VA) and three Midwestern (IA, MN, OH) states, as compared to RABapp™, and the USDA, 2022 Census of Agriculture data (USDA, 2024). Farm capacity was used as a surrogate for population size for RABapp™ farms

| | Barns | | Farms | | | Population size* | | | |
|---|---|---|---|---|---|---|---|---|---|
| State | RABapp™ | Predicted | RABapp™ | USDA | Predicted | RABapp™ | USDA | Predicted | % Diff [1] |
| IA | 3,375 | 15,403 | 1,501 | 5,419 | 7,998 | 6.06 M | 24.6 M | 27.4 M | + 11 % |
| MN | 974 | 10,326 | 259 | 3,180 | 6,178 | 0.97 M | 9.2 M | 17.7 M | + 92 % |
| OH | 536 | 3,194 | 210 | 2,137 | 1,981 | 0.52 M | 2.45 M | 5.74 M | + 135 % |
| NC | 8,165 | 13,894 | 1,721 | 2,492 | 4,063 | 6.91 M | 8.19 M | 19.6 M | + 140 % |
| SC | 195 | 1,746 | 33 | 370 | 928 | 0.18 M | 0.01 M | 2.9 M | + 28,900 % |
| VA | 332 | 1,017 | 48 | 799 | 439 | 0.29 M | 0.06 M | 1.6 M | + 2,600 % |

* Population size estimates are rounded to the nearest hundred thousand. "M" denotes millions of pigs

[1] Percent difference (% Diff) represents the percent difference between predicted and USDA numbers and was calculated as follows: $Percent\ difference = \frac{Predicted - USDA}{USDA} \times 100$

## 4. Discussion

This study presents a four-stage machine learning framework that integrates semantic segmentation, classification, and regression to predict the location, production type, and population size of commercial swine farms across two regions in the U.S. Our results demonstrated significant differences in the spatial and demographic information currently available, predicting 63% more farms than the USDA in North Carolina (Predicted: 4,063 vs. USDA: 2,492), 151% more in South Carolina (Predicted: 928 vs. USDA: 370), and 45% fewer in Virginia (Predicted: 439 vs. USDA: 799) (USDA NASS, 2024; USDA, 2024). In the Midwestern region, the number of predicted farms exceeded USDA estimates by 54% in Iowa

(Predicted: 7,998 vs. USDA: 5,419), 94% in Minnesota (Predicted: 6,178 vs. USDA: 3,180), and showed a modest underestimation of 7% in Ohio (Predicted: 1,981 vs. USDA: 2,137). The semantic segmentation model achieved an F2-score of 92% and a mean IoU of 76%, indicating that combining machine learning with aerial imagery could be an effective approach for detecting commercial swine farms. However, spectral and structural similarities between swine farms and other features (e.g., poultry barns, warehouses, mobile homes) resulted in 74% to 96% of the predicted polygons being flagged as false positives by our Random Forest-based filtering model. Furthermore, our production type classification model correlated with distributions in the RABapp™ dataset, the most comprehensive dataset of U.S. swine farms, highlighting the model's ability to capture realistic farm-type distributions. While our total pig population estimates exceed current USDA reports, these numbers suggest that the U.S. swine population could surpass the USDA's reported number of farms and animals.

    Our findings build on previous efforts to map livestock operations using remotely sensed imagery and machine learning techniques (Saha et al., 2025; Tulbure et al., 2024; Robinson et al., 2022; Montefiore et al., 2022; Patyk et al., 2020; Maroney et al., 2020; Handan-Nader and Ho, 2019; Burdett et al., 2015). The semantic segmentation model identified swine barns from high-resolution aerial images; however, it also produced false positives due to spectral and structural similarities between swine barns and other agricultural (e.g., poultry barns, cattle barns, and farm storage sheds), industrial (e.g., warehouses and greenhouses), or residential buildings (e.g., mobile homes and large garages). Visual inspection revealed an association between false-positive detections and certain NLCD land cover classes, particularly developed and agricultural areas. To address this challenge, we implemented a Random Forest-based filtering model, similar to the approach applied by Tulbure et al. (2024), which reduced over

50% of false positives from poultry barn predictions for the U.S (Robinson et al., 2022). Similarly, our filtering model eliminated between 74% to 96% of potential false positives, substantially refining the predicted dataset. In addition to false positives, we observed that the predicted barn footprints often extended beyond the building boundaries, leading to inflated estimates of barn area. While the prediction contained the feature of interest, the additional area included in the prediction may have important implications for downstream results, such as production type classification and population estimation.

Identifying the production type of commercial swine farms is essential for modeling disease dynamics (Cardenas et al., 2024; Galvis and Machado, 2024; Sanchez et al., 2023; Sykes et al., 2023; Galvis et al., 2022; Gilbertson et al., 2022; Campler et al., 2021). While many studies have focused on locating farms and estimating their population sizes, none to our knowledge have attempted to classify them into production types (Saha et al., 2025; Robinson et al., 2022; Handan-Nader et al., 2021; Burdett et al., 2015; Patyk et al., 2020; Martin et al., 2015). This distinction is important because each production type plays a distinct role in the swine industry's movement network, influencing the direction and frequency of animal movements, as well as the associated disease transmission dynamics (Cardenas et al., 2024; Sykes et al., 2023; Galvis et al., 2022, 2021). Classifying production type, however, is challenging due to structural similarities among barn types. Our classification of production types achieved an overall accuracy of 87%. Misclassifications were primarily observed between nursery and finisher farms, which often share similar architectural features at aerial scales, whereas sow farms were more consistently identified. The model successfully leveraged barn-level heuristics such as barn length, width, area, aspect ratio, and number of barns in a farm to capture differences between sow, nursery, and finisher farms. Among these, barn length, width, and area were important predictors in the classification model, which have also been noted by other studies as key features for distinguishing livestock facilities (Tulbure et al., 2024; Robinson et al., 2022). The predicted production type proportions were correlated with RABapp™ data (RABapp, 2024), lending support to the reliability of our model's output. However, performance was poorest for boar stud farms, which was expected given their small sample size, limited distribution, and lack of consistent structural characteristics. These farms tend to be geographically isolated, single-barn operations due to biosecurity requirements and were often misclassified as other production types, including finisher and nursery farms, because of their structural ambiguity and underrepresentation in the training data.

The performance of our population estimation model was fair ($R^2 = 0.77$); however, estimates were subject to compounding errors from upstream workflow stages, including residual false positives and misclassification of production types. The model performed better for small to medium-sized farms and exhibited greater error when predicting the populations of larger operations. Performance also varied by production type, with nursery farms showing the highest agreement ($R^2 = 0.82$), followed by finisher ($R^2 = 0.77$) and sow farms ($R^2 = 0.56$). While population-size predictions for boar stud farms performed poorly overall, the predicted values still followed the general population trend. This discrepancy is likely attributable to the distinct configuration of boar stud facilities, which typically allocate substantially more space per animal than other production systems, resulting in differences in barn-to-population scaling relationships. In contrast to previous efforts that rely on aggregated estimates from the Census of Agriculture (CoA), which are limited by survey nonresponse, temporal lags, and data suppression in sparsely populated areas, our model was trained on the animal capacity reported by production companies (Fleming et al., 2025; RABapp, 2024). This approach enables more fine-scale and up-to-date estimates of swine populations. Although our total population estimates exceeded those reported by the USDA, these results may reflect a more complete accounting of undocumented or underreported farms and suggest that the U.S. swine population could be higher than current federal statistics indicate.

Our approach offers a scalable alternative to survey-based methods by addressing persistent gaps in the spatial and demographic representation of commercial swine farms in the U.S. The proposed framework not only detects farm locations but also classifies their production types and estimates swine populations, thereby enhancing the granularity and operational relevance of available data. These capabilities have important implications for disease monitoring and surveillance, enabling the development of spatially informed epidemiological models (Fleming et al., 2025; Brandon H. Hayes et al., 2024; B. H. Hayes et al., 2024; Galvis and Machado, 2024; Brandon H. Hayes et al., 2023; Cardenas et al., 2022; Gilbertson et al., 2022; McBride et al., 2021; Makau, Alkhamis et al., 2021; Pudenz et al., 2019; Rossi et al., 2017). By improving our understanding of where farms are, what types of operations they represent, and how many animals they house, this approach supports risk assessments, resource allocation, and intervention strategies in the face of emerging and transboundary diseases (Galvis and Machado, 2024; Dupas et al., 2024; Sykes et al., 2023; Moraes et al., 2023; Pepin et al., 2022; Gilbert et al., 2022; Gilbertson et al., 2022; Campler et al., 2021; Kao et al., 2006; Robinson et al., 2022).

## 5. Limitations and final remarks

This study predicted swine farm locations and farm-level demographic information; however, limitations should be acknowledged. First, the presence of a detected barn does not confirm that the farm is currently active. Our predictions are based on aerial imagery collected in 2022 and 2023 (USDA, 2022), which captures recently constructed structures but does not confirm the farm's operational status. Second, some farms may have been constructed after the imagery was acquired; therefore, they would not be captured in our predictions.

Class imbalances between production types posed a challenge for training our models to classify mixed or less common production types. Many swine farms specialize in more than one aspect of the production cycle (e.g., farrow-to-finish or wean-to-finish systems), making it difficult to assign a single, representative category. To help mitigate this issue, production types were aggregated into four broad categories (sow, nursery, finisher, and boar stud). However, this generalization may still result in misclassification, particularly for farms that operate across multiple stages of the production cycle. Also, our model was developed specifically to detect commercial-scale operations and is not designed to identify backyard farms. These smaller operations, which have grown in recent years (USDA, 2024, 2019), often fall outside the detection range due to their varied or less distinguishable structural features. Additionally, our aggregation of barns into farm units using a 500-meter proximity threshold may have inadvertently grouped distinct operations that are located near one another or, conversely, separated barns belonging to the same producer if they were distributed across non-contiguous sites. While this approach provides a consistent method for approximating farm boundaries, it introduces uncertainty when ownership or operational control spans multiple spatially dispersed facilities.

Our population size estimates reflect the farm's pig capacity rather than the current number of animals. Swine populations fluctuate throughout production cycles, and without temporal data, these estimates should be interpreted as the maximum potential occupancy rather than actual headcounts (Fleming et al., 2025; Cardenas et al., 2024; RABapp, 2024). The cumulative nature of errors across workflow stages, such as false positives from the segmentation model or misclassified production types, may also affect the accuracy of downstream population estimates. We observed that predicted barn footprints occasionally extended beyond building boundaries, resulting in inflated area estimates (Figure 3 and Supplementary Material Figures S5 and S7). While the Random Forest-based filtering model substantially reduced false positives and improved overall precision, compounding errors remain and should be considered when interpreting the final outputs.

To account for regional variation in production systems that are likely to impact our model performance, such as land cover and waste management, we developed separate models for the Southeastern and Midwestern U.S. This approach enabled the workflow to capture regional characteristics. While this strategy addressed geographic variability within our study area, further testing is needed to evaluate scalability to other regions.

## 6. Conclusion

Results from our four-stage machine learning framework represent a significant advancement in the detection and classification of commercial swine barns using high-resolution aerial imagery. By integrating deep learning and Random Forest models, our approach distinguished swine barns from other structures across diverse landscapes. Subsequent Random Forest classification and regression models enabled the prediction of production types and farm-level population sizes. This comprehensive, data-driven approach serves as a valuable complement to existing monitoring systems, supporting more accurate modeling of animal health risks and enhancing the capacity for targeted interventions across the U.S. swine sector.

**Funding**

This project was funded by the Center for Geospatial Analytics and the College of Veterinary Medicine at North Carolina State University. This work was also supported by the Foundation for Food & Agriculture Research (FFAR) award number FF-NIA21-0000000064.

**Author contributions**

Felipe E. Sanchez: Conceptualization, Visualization, Validation, Software, Investigation, Methodology, Writing – original draft, and Writing – review & editing. Thomas A. Lake: Conceptualization, Validation, Methodology, Writing – review & editing. Jason A. Galvis: Conceptualization, Methodology, Writing – review & editing. Chris Jones: Conceptualization, Supervision, Resources, Funding acquisition, Investigation, Methodology, and Writing – review & editing. Gustavo Machado: Conceptualization, Investigation, Supervision, Writing – review & editing, Project administration; Resources, Funding acquisition.

Appendix

**Predicting the spatial distribution and demographics of commercial swine farms in the United States.**

**Section S1. Reclassification of RABapp™ commercial swine production types**

Commercial swine farms in the RABapp™ dataset were grouped into four primary production types—finisher, sow, nursery, and boar stud—based on company-reported designations. The following descriptions outline the functional role of each category and summarize the reclassification scheme used in Supplementary Material Table S1. Sow premises encompass locations equipped for breeding, gestation, or farrowing activities. A gilt development unit (GDU) is a farm or barn dedicated to the raising of replacement female pigs (gilts) that have not given birth to a litter before entering the main sow breeding herd. Nursery premises are designated for raising piglets from approximately three weeks to about ten weeks of age. Finisher premises focus on the growth and development of pigs from around ten weeks of age until they reach market weight, typically between five and six months. Boar studs are premises housing male pigs of reproductive age.

1 **Supplementary Material Table S1.** Reclassification of commercial swine production types
2 based on the provided production type by the company.

| Reclassified production types | Production types provided by the companies |
|---|---|
| **Sow** | "GDU", "Developer", "Gilt Finishing", "Gilt Isolation", "Isolation", "Gilt Growout", "Sow; Finishing", "Gilt", "Gilt Breeder", "Isolation; Sow", "GDU Finisher", "GDU Nursery", "Developer" |
| **Nursery** | "Nursery" |
| **Finisher** | "Wean to Finish", "Farrow to Finish", "Research", "Nursery; Finisher", "Finish" |
| **Boar stud** | "Boar", "Boar stud" |

## Section S2. Deriving barn polygons from RABapp™ data

Using line of separation (LOS) features from 5,160 commercial swine farms in the RABapp™ database (Supplementary Figure S1), we derived individual barn polygons to support multiple components of the modeling workflow, including the creation of training labels for the semantic segmentation model (Section 2.4 and Supplementary Section S3). A centroid was placed within each barn structure, and linear features were manually digitized to delineate individual barns from adjacent walkways and nearby structures (Supplementary Figure S2). These lines were used to subdivide each LOS polygon into discrete barn polygons. The resulting geometries were rasterized to generate binary mask tiles, which were then aligned with their corresponding NAIP image tiles. Within each mask tile, pixels were labeled as either "barn" (1) or "background" (0) (Figure 1-b).

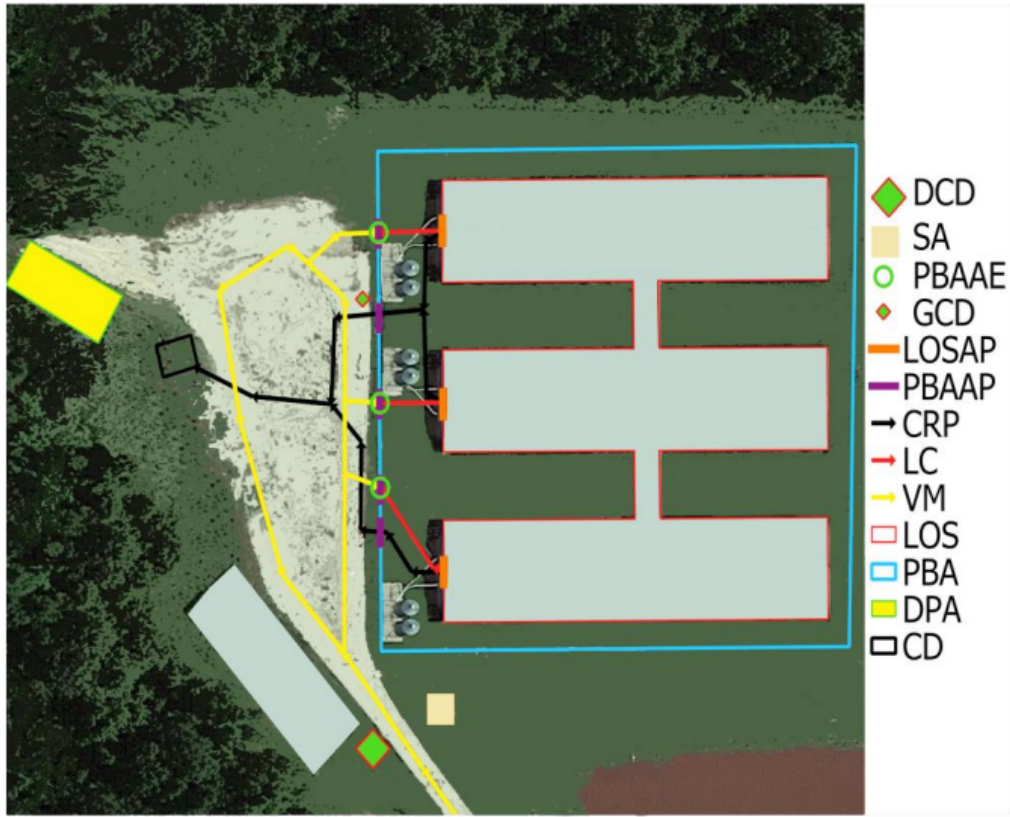

16  **Supplementary Material Figure S1**. **Example of a map within the RABapp™ database.**

17  Example of a swine farm from the RABapp™ database illustrating on-farm biosecurity features.

18  The line of separation (LOS) delineates the boundary enclosing animal housing and sanitized

19  zones for personnel and equipment (RABapp, 2024).

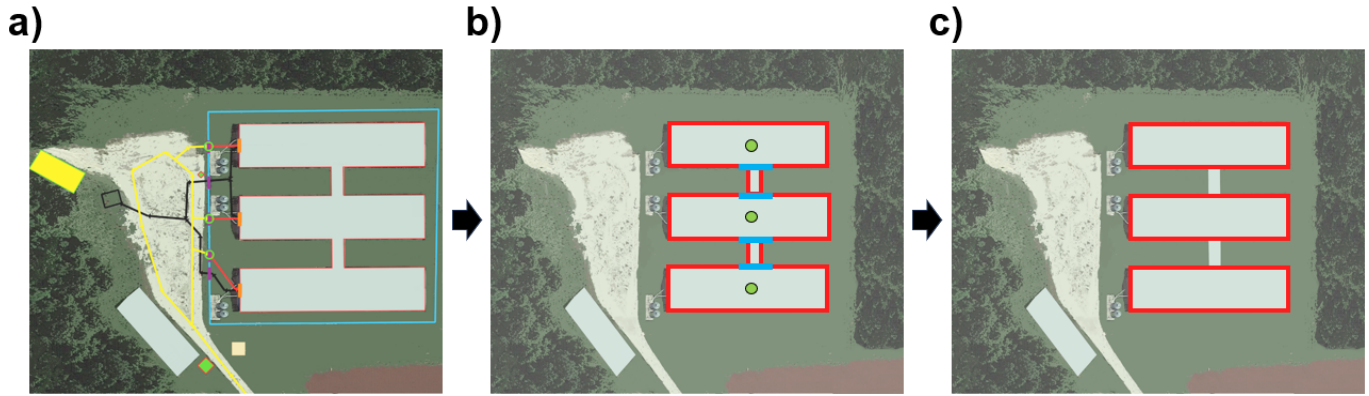

**Supplementary Material Figure S2. Illustration of the barn division process.** a) Example of a farm within the RABapp™ database with on-farm biosecurity features (Fleming et al., 2025; RABapp, 2024). b) Centroids (green dots) and barn divisions (blue lines) were added to split the single line of separation (LOS) polygon feature into separate barns. c) Separate barns were produced after splitting the LOS feature and were included in our ground truth dataset.

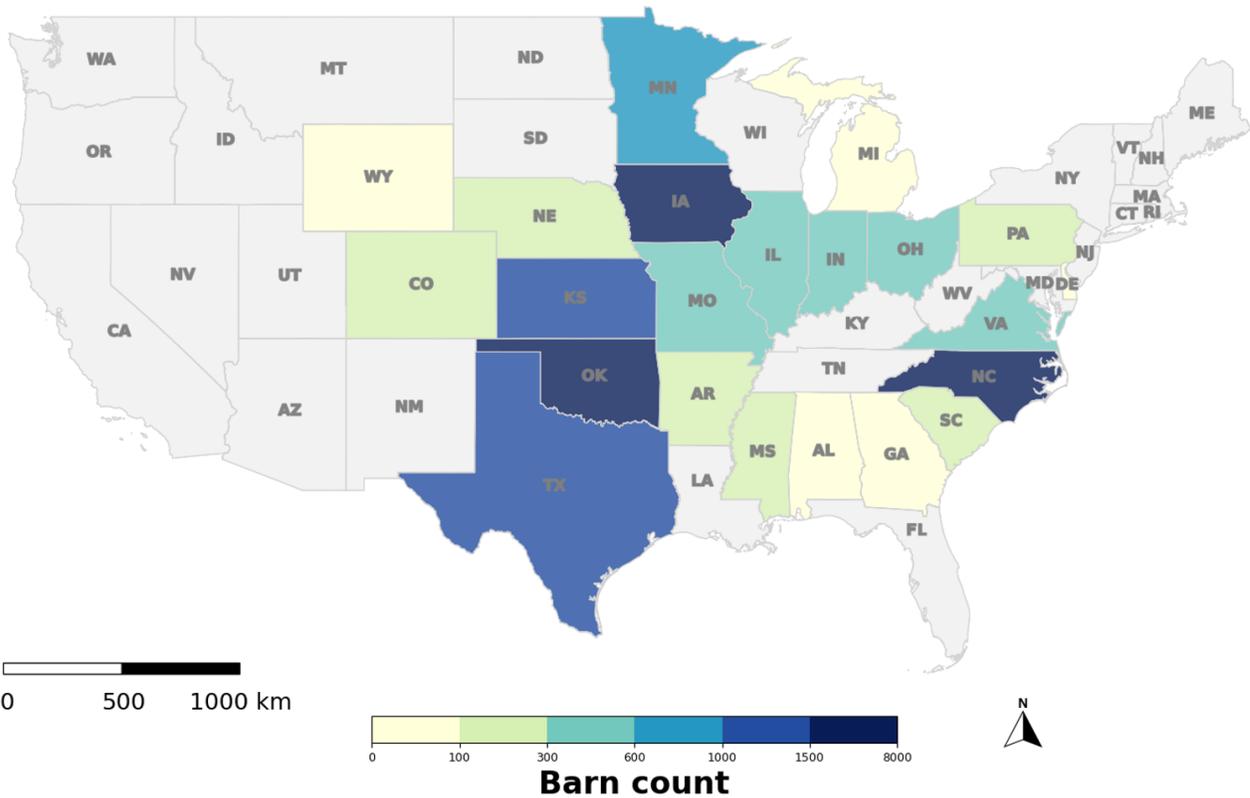

28   **Supplementary Material Figure S3**. Map of the contiguous U.S. showing the density of barns

29   per state in the RABapp™ database (Fleming et al., 2025).

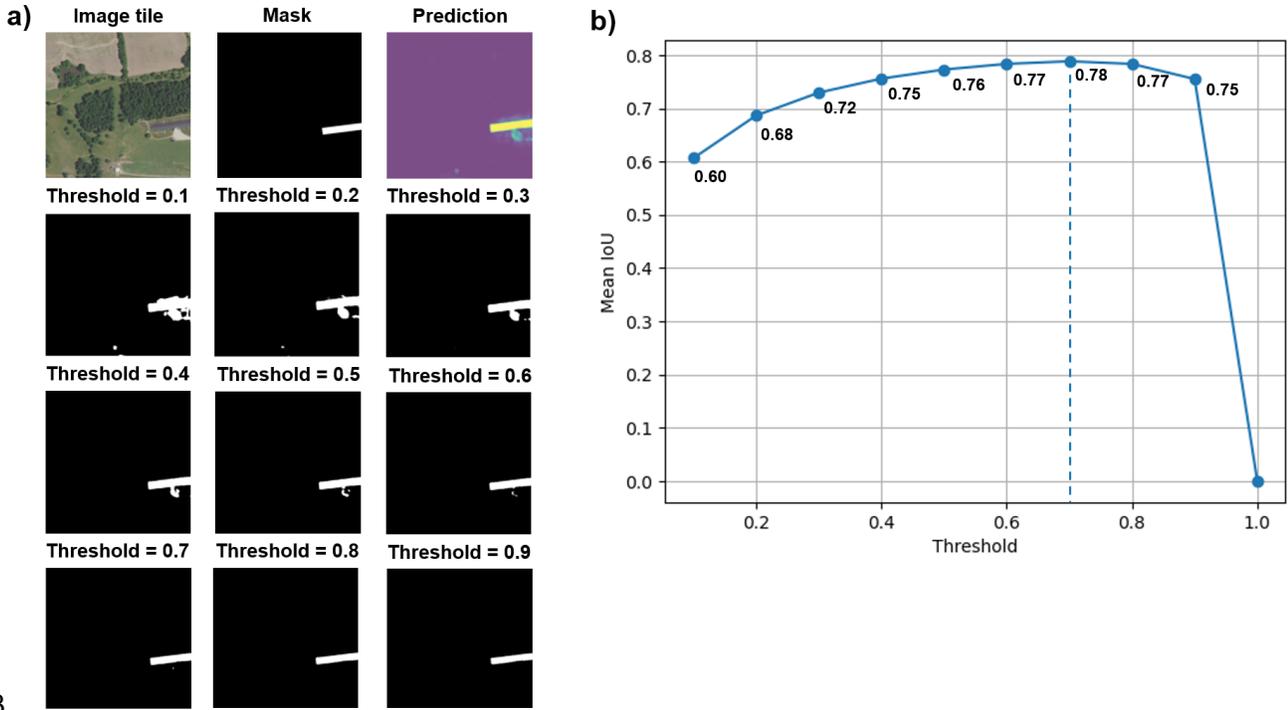

31   **Supplementary Material Figure S4.** a) Threshold analysis for the semantic segmentation

32   of a swine barn. The first row shows the original 512 × 512 meter NAIP RGB image tile

33   (left), the corresponding binary ground-truth mask where barns are labeled in white (center),

34   and the raw model output before thresholding (right). Subsequent rows display binary

35   predictions [0, 1] for thresholds ranging from 0.1 to 0.9 in 0.1 increments. As the threshold

36   increases, predictions become more conservative, with fewer pixels classified as barn

37   increments. b) Mean Intersection over Union (IoU) values computed across validation tiles

38   for each threshold. A threshold of 0.7 (dashed line) yielded the highest average IoU (0.78),

39   indicating optimal agreement between predicted and ground-truth barn regions.

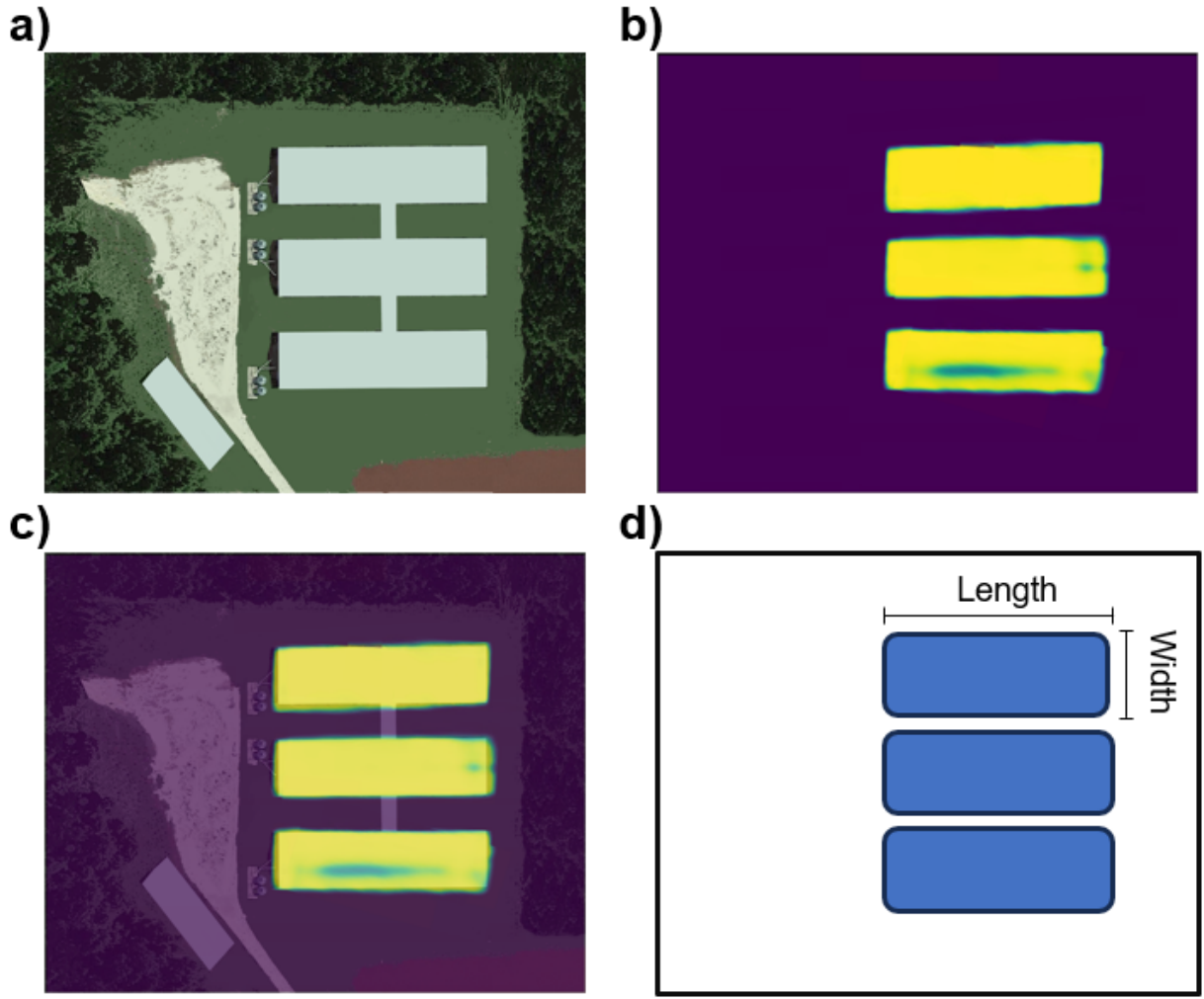

**Supplementary Material Figure S5**. a) Example image of a swine barn in the Rabapp™ database. b) Probability heatmap showing the likelihood of each pixel belonging to a barn structure (yellow = high probability, purple = low). c) Overlay of the probability predictions on the original aerial image to visualize alignment with actual barns. d) Final barn polygons extracted from the probability map using an 8-connected pixel grouping approach, with length and width measurements derived from the polygon geometry.

47    **Section S3. Semantic segmentation performance metrics**

48    Metrics used to evaluate the semantic segmentation model's performance were derived from the

49    confusion matrix, which summarizes true positives, false positives, true negatives, and false

50    negatives ($FN$) obtained from predictions on an out-of-sample testing dataset. These metrics

51    included overall accuracy, precision, recall (sensitivity), specificity, mean Intersection over

52    Union (IoU), and the F2-score. Formulas for each metric are detailed below. Overall accuracy

53    measures the proportion of correctly identified pixels and is defined as:

$$(TP + TN) / (TP + TN + FP + FN) \quad (1)$$

55    Precision quantifies the proportion of correctly predicted barn pixels out of all pixels predicted

56    as barns:

$$TP / (TP + FP) \quad (2)$$

59    Recall (sensitivity) evaluates the model's ability to identify all true positives and is calculated

60    as:

$$TP / (TP + FN) \quad (3)$$

62    Specificity assesses the model's ability to identify true negatives correctly and is calculated as:

$$TN / (TN + FP) \quad (4)$$

65    Mean Intersection over Union (IoU) quantifies the overlap between predicted and ground-truth

66    barn pixels relative to their combined area:

$$(TP) / (TP + FP + FN) \quad (5)$$

68    F1-score is a weighted harmonic mean of the precision and recall:

$$\text{F1-score} = \frac{2 \cdot precision \cdot recall}{precision + recall} \quad (6)$$

F2-score is a weighted harmonic mean of the precision and recall that weights recall higher than precision:

$$\text{F2-score} = \frac{(1+2^2) \cdot (precision \cdot recall)}{(2^2 \cdot precision) + recall} \tag{7}$$

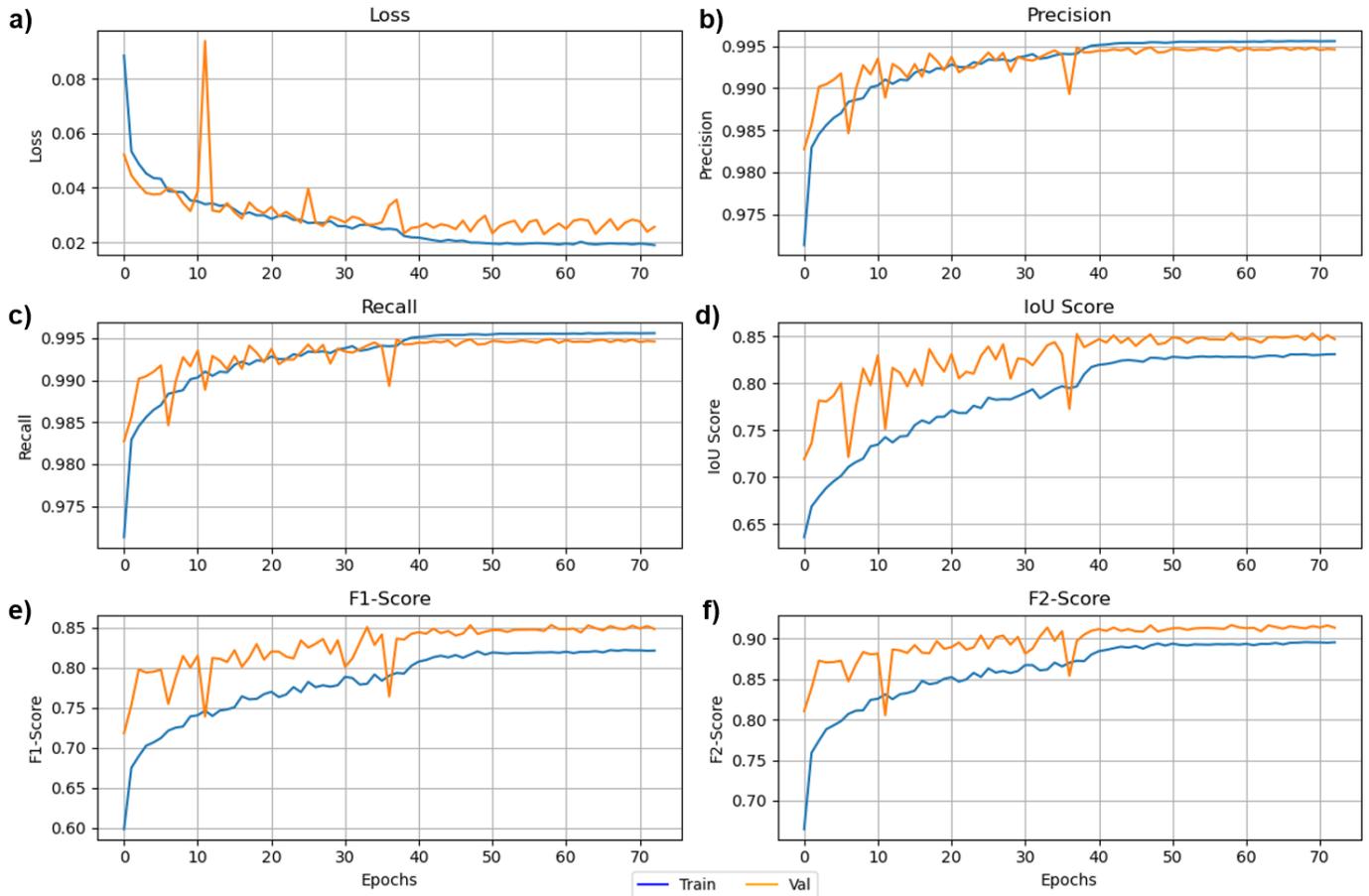

**Supplementary Material Figure S6. Performance metrics for the U-Net semantic segmentation model.** Each plot compares the training and validation performance across 73 epochs for a) Binary cross-entropy loss, b) Precision, c) Recall, d) Intersection over Union (IoU) score, e) F1-score, and f) F2-score.

The confusion matrix showed that most misclassifications were false positives (9.96 million pixels, or 0.30% of the test area) rather than false negatives (1.59 million pixels, or 0.05% of the test area). In contrast, the model correctly identified 3.33 billion true negatives (99.35% of the test area) and 37.4 million true positives (1.12% of the test area), suggesting that while the model reliably detected barns, it also misidentified some non-barn features as barns.

**Section S4. Random forest barn feature filtering**

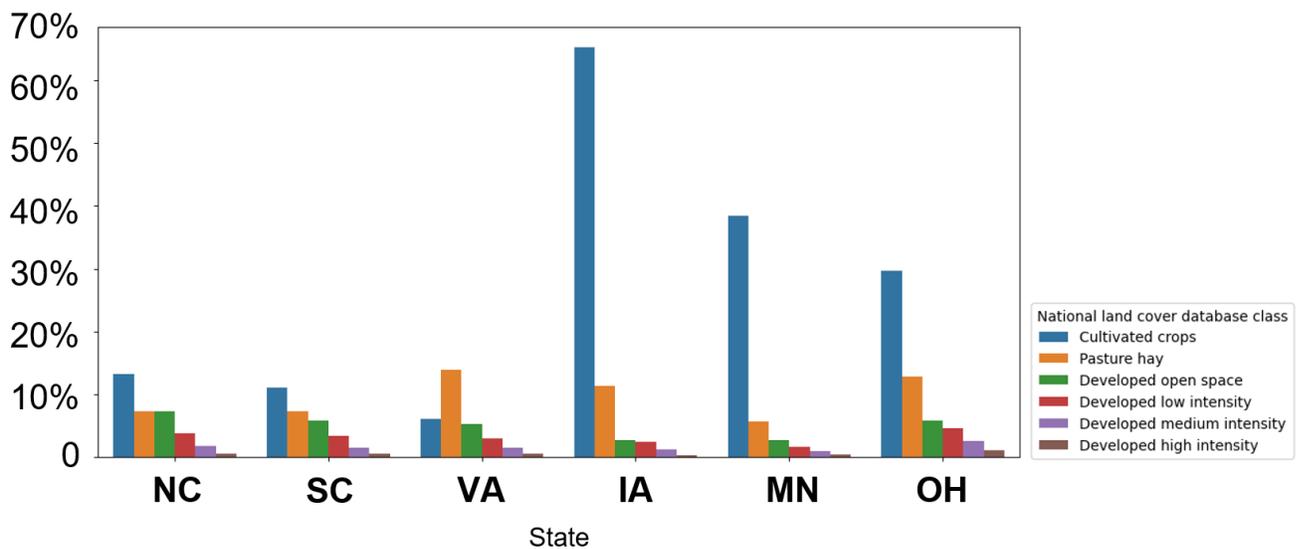

**Supplementary Material Figure S7. National Land Cover Database (NLCD) land cover class proportions by state.** Land cover proportions were calculated using all NLCD classes for each state; however, only the six classes most commonly found within a 500 m buffer around swine farms are shown here: Cultivated Crops, Pasture/Hay, and four Developed categories.

**Supplementary Table S2.** Distribution of training and testing samples used in 5-fold spatial cross-validation for the Random Forest barn filtering model of the Southeastern region, including the following states: NC, SC, VA.

| Fold | Training | Testing | Training | | | Testing | | |
|---|---|---|---|---|---|---|---|---|
| | | | Barns | False positives | Total | Barns | False positives | Total |
| 1 | 2, 3, 4, 5 | 1 | 7,504 | 16,139 | 23,643 | 1,191 | 6,371 | 7,562 |
| 2 | 1, 3, 4, 5 | 2 | 7,297 | 16,010 | 23,307 | 1,398 | 6,500 | 7,898 |
| 3 | 1, 2, 4, 5 | 3 | 7,293 | 18,947 | 26,240 | 1,402 | 3,563 | 4,965 |
| 4 | 1, 2, 3, 5 | 4 | 6,316 | 21,162 | 27,478 | 2,379 | 1,348 | 3,727 |
| 5 | 1, 2, 3, 4 | 5 | 6,370 | 17,782 | 24,152 | 2,325 | 4,729 | 7,053 |

94

95 **Supplementary Table S3.** Distribution of training and testing samples used in 5-fold spatial
96 cross-validation for the Random Forest barn filtering model of the Midwestern region, including
97 the following states: IA, MN, OH.

| Fold | Training | Testing | Training | | | Testing | | |
|---|---|---|---|---|---|---|---|---|
| | | | Barns | False positives | Total | Barns | False positives | Total |
| 1 | 2, 3, 4, 5 | 1 | 3,712 | 10,990 | 14,702 | 1,173 | 1,324 | 2,497 |
| 2 | 1, 3, 4, 5 | 2 | 3,807 | 8,944 | 12,751 | 1,078 | 3,370 | 4,448 |
| 3 | 1, 2, 4, 5 | 3 | 3,816 | 11,665 | 15,481 | 1,069 | 649 | 1,718 |
| 4 | 1, 2, 3, 5 | 4 | 4,047 | 9,834 | 13,881 | 838 | 2,480 | 3,318 |
| 5 | 1, 2, 3, 4 | 5 | 4,158 | 7,823 | 11,981 | 727 | 4,491 | 5,218 |

98

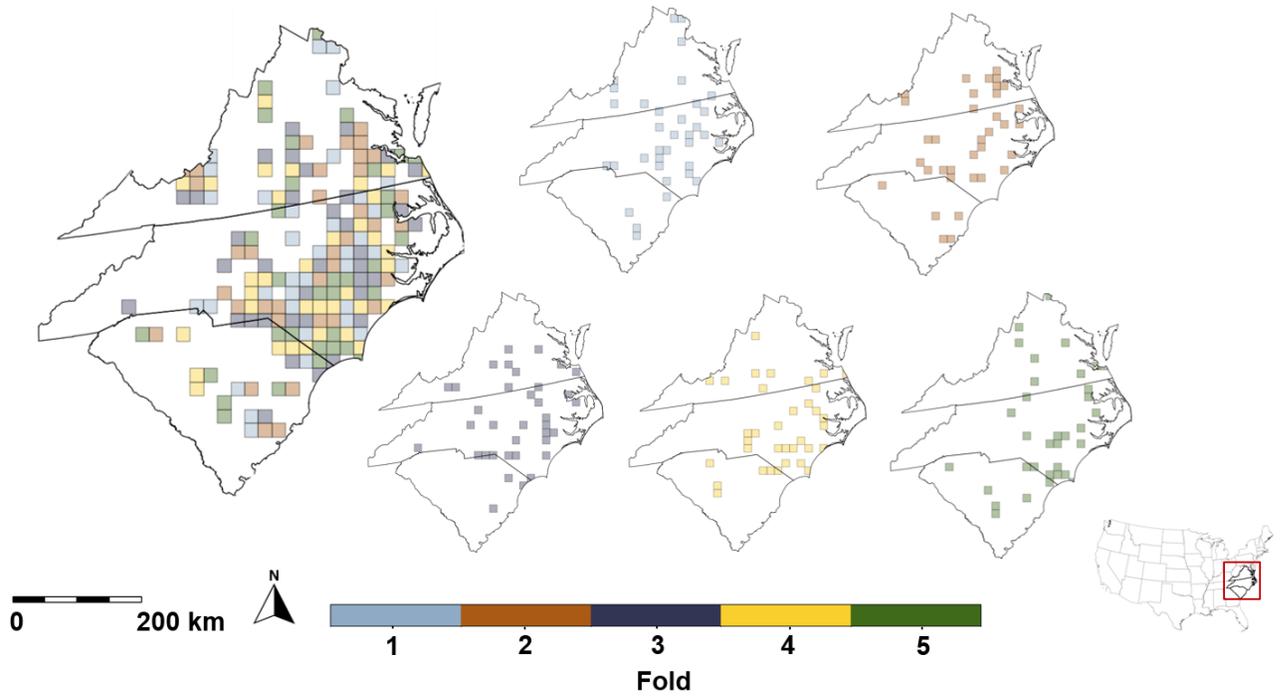

**Supplementary Figure S8. 5-fold spatial cross-validation across Southeastern and Midwestern States.** Spatial allocation of 25 × 25 km blocks used in the 5-fold spatial cross-validation procedure for the Southeastern (NC, SC, VA) region. Each fold consisted of non-overlapping spatial blocks containing either barns/or false positive polygons.

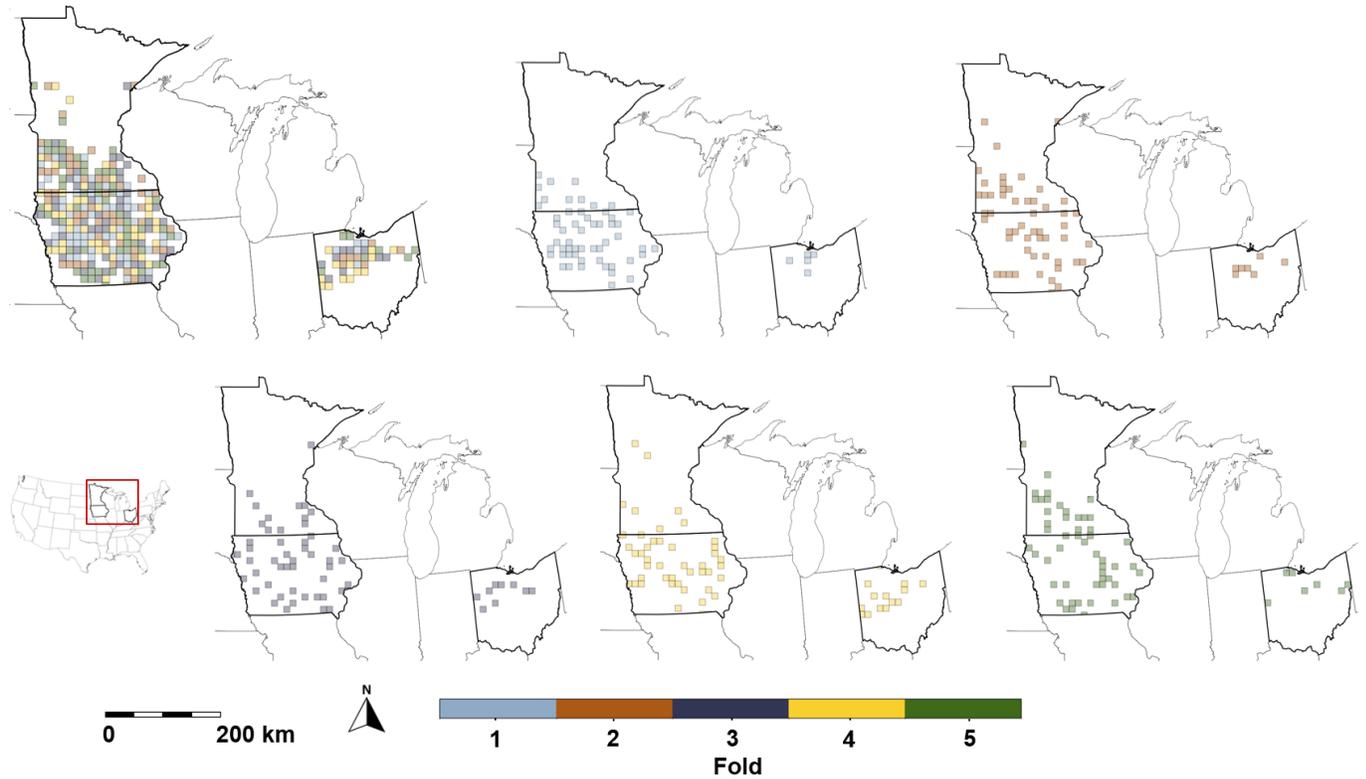

**Supplementary Figure S9. 5-fold spatial cross-validation across the Midwestern States.** Spatial allocation of 25 × 25 km blocks used in the 5-fold spatial cross-validation procedure for the Midwestern (IA, MN, OH) regions. Each fold consisted of non-overlapping spatial blocks containing either barns/or false positive polygons.

**Supplementary Table S4.** Performance metrics of the Random Forest barn filtering model of the Southeastern region (NC, SC, VA)

| Fold | Buffer (m) | Training | Testing | Accuracy | Precision | Recall | F1 Score |
|---|---|---|---|---|---|---|---|
| 1 | 500 | 2, 3, 4, 5 | 1 | 0.948 | 0.995 | 0.999 | 0.997 |
| 2 | 500 | 1, 3, 4, 5 | 2 | 0.960 | 0.950 | 1.000 | 0.974 |
| 3 | 500 | 1, 2, 4, 5 | 3 | 0.999 | 1.000 | 0.996 | 0.988 |
| 4 | 500 | 1, 2, 3, 5 | 4 | 0.955 | 0.999 | 0.991 | 0.995 |
| 5 | 500 | 1, 2, 3, 4 | 5 | 0.985 | 1.000 | 0.983 | 0.991 |
| 1 | 1000 | 2, 3, 4, 5 | 1 | 0.997 | 0.996 | 0.996 | 0.996 |
| 2 | 1000 | 1, 3, 4, 5 | 2 | 0.993 | 0.964 | 1.000 | 0.982 |
| 3 | 1000 | 1, 2, 4, 5 | 3 | 0.998 | 1.000 | 0.989 | 0.994 |
| 4 | 1000 | 1, 2, 3, 5 | 4 | 0.998 | 1.000 | 0.996 | 0.998 |
| 5 | 1000 | 1, 2, 3, 4 | 5 | 0.992 | 0.999 | 0.975 | 0.987 |
| 1 | 5000 | 2, 3, 4, 5 | 1 | 0.998 | 0.995 | 0.999 | 0.997 |
| 2 | 5000 | 1, 3, 4, 5 | 2 | 0.990 | 0.953 | 0.999 | 0.995 |
| 3 | 5000 | 1, 2, 4, 5 | 3 | 0.999 | 1.00 | 0.994 | 0.997 |
| 4 | 5000 | 1, 2, 3, 5 | 4 | 0.995 | 0.999 | 0.991 | 0.995 |
| 5 | 5000 | 1, 2, 3, 4 | 5 | 0.995 | 1.00 | 0.983 | 0.991 |

112 **Supplementary Table S5.** Performance metrics for the Random Forest barn filtering model of
113 the Midwestern region (IA, MN, OH)

| Fold | Buffer (m) | Training | Testing | Accuracy | Precision | Recall | F1 Score |
|---|---|---|---|---|---|---|---|
| 1 | 500 | 2, 3, 4, 5 | 1 | 0.987 | 0.979 | 1.000 | 0.989 |
| 2 | 500 | 1, 3, 4, 5 | 2 | 0.920 | 0.798 | 1.000 | 0.888 |
| 3 | 500 | 1, 2, 4, 5 | 3 | 0.977 | 1.000 | 0.976 | 0.988 |
| 4 | 500 | 1, 2, 3, 5 | 4 | 0.981 | 0.992 | 0.984 | 0.988 |
| 5 | 500 | 1, 2, 3, 4 | 5 | 0.904 | 0.882 | 0.956 | 0.918 |
| 1 | 1000 | 2, 3, 4, 5 | 1 | 0.989 | 0.994 | 1.000 | 0.997 |
| 2 | 1000 | 1, 3, 4, 5 | 2 | 0.994 | 0.862 | 0.993 | 0.923 |
| 3 | 1000 | 1, 2, 4, 5 | 3 | 0.995 | 1.000 | 0.964 | 0.982 |
| 4 | 1000 | 1, 2, 3, 5 | 4 | 0.999 | 0.992 | 0.992 | 0.992 |
| 5 | 1000 | 1, 2, 3, 4 | 5 | 0.984 | 0.907 | 0.963 | 0.935 |
| 1 | 5000 | 2, 3, 4, 5 | 1 | 0.974 | 0.870 | 0.984 | 0.924 |
| 2 | 5000 | 1, 3, 4, 5 | 2 | 0.937 | 0.945 | 0.963 | 0.954 |
| 3 | 5000 | 1, 2, 4, 5 | 3 | 0.999 | 0.975 | 0.992 | 0.983 |
| 4 | 5000 | 1, 2, 3, 5 | 4 | 0.998 | 0.996 | 0.984 | 0.990 |
| 5 | 5000 | 1, 2, 3, 4 | 5 | 0.954 | 0.720 | 0.974 | 0.828 |

**Supplementary Table S6.** Number of predicted barn polygons and percent reduction by Southeastern states and buffer distance (500 m, 1000 m, 5000 m).

| State | Buffer (m) | Predicted Polygons | Voted Barns | Reduction (%) |
|---|---|---|---|---|
| NC | 500 | 111,135 | 38,966 | 65 |
| NC | 1000 | 111,135 | 28,648 | 74 |
| NC | 5000 | 111,135 | 36,576 | 67 |
| SC | 500 | 37,264 | 7,616 | 80 |
| SC | 1000 | 37,264 | 4,138 | 89 |
| SC | 5000 | 37,264 | 6,667 | 82 |
| VA | 500 | 46,075 | 3,019 | 94 |
| VA | 1000 | 46,075 | 1,942 | 96 |
| VA | 5000 | 46,075 | 2,455 | 95 |

119    **Supplementary Table S7.** Number of predicted barn polygons and percent reduction by
120    Midwestern states and buffer distance (500 m, 1000 m, 5000 m).

| State | Buffer (m) | Predicted Polygons | Voted Barns | Reduction (%) |
|---|---|---|---|---|
| IA | 500 | 168,866 | 45,212 | 73 |
| IA | 1000 | 168,866 | 41,669 | 75 |
| IA | 5000 | 168,866 | 30,286 | 82 |
| MN | 500 | 165,714 | 45,551 | 73 |
| MN | 1000 | 165,714 | 40,378 | 76 |
| MN | 5000 | 165,714 | 25,033 | 85 |
| OH | 500 | 190,832 | 27,934 | 85 |
| OH | 1000 | 190,832 | 19,206 | 90 |
| OH | 5000 | 190,832 | 7,350 | 96 |

121

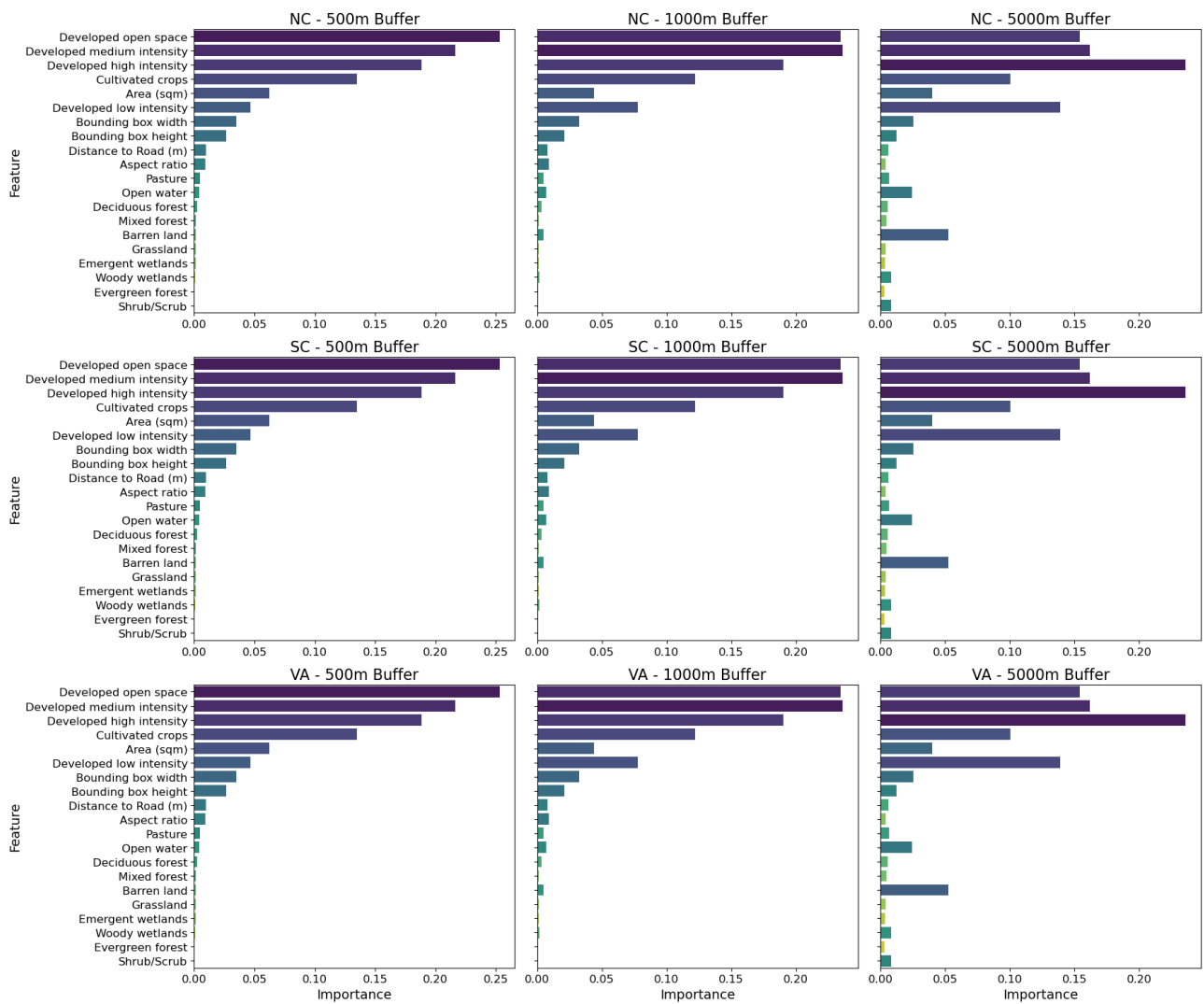

**Supplementary Material Figure S10. Average feature importance (Gini index) by state (rows) and NLCD buffer distances (columns) for the Southeastern region.** Each bar plot shows the top-ranked predictors based on mean importance across five spatial cross-validation folds for North Carolina (NC), South Carolina (SC), and Virginia (VA). Columns represent buffer distances of 500 m, 1000 m, and 5000 m used to extract proportions of surrounding NLCD land cover classes.

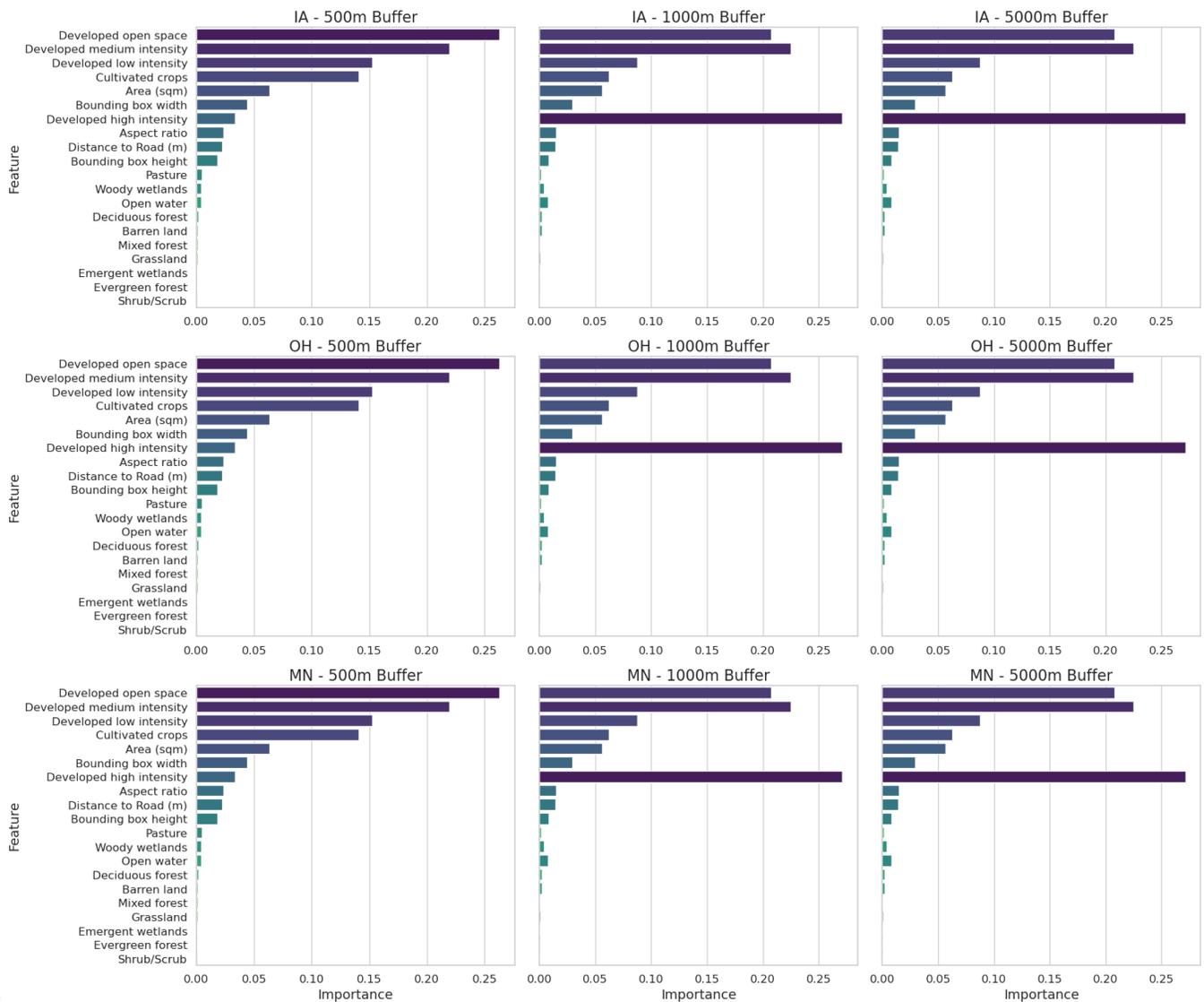

**Supplementary Material Figure S11. Average feature importance (Gini index) by state (rows) and NLCD buffer distances (columns) for the Midwestern region.** Each bar plot shows the top-ranked predictors based on mean importance across five spatial cross-validation folds for Iowa (IA), Ohio (OH), and Minnesota (MN). Columns represent buffer distances of 500 m, 1000 m, and 5000 m used to extract proportions of surrounding NLCD land cover classes.

137 **Supplementary Table S8.** Number of predicted barn polygons that intersect at least one OSM
138 building with the listed tag in the Southeastern states (NC, SC, VA).

| OpenStreetMap© building tag | Count |
| --- | --- |
| yes | 6000 |
| farm_auxiliary | 1588 |
| sty | 664 |
| school | 453 |
| farm | 319 |
| industrial | 236 |
| chicken shed | 197 |
| commercial | 181 |
| chicken_shed | 161 |
| warehouse | 145 |
| barn | 106 |
| data_center | 93 |
| hangar | 83 |
| greenhouse | 72 |
| retail | 68 |
| terrace | 63 |
| garages | 47 |
| roof | 45 |
| apartments | 43 |
| house | 31 |
| parking | 30 |
| residential | 29 |
| stable | 14 |
| church | 14 |
| grandstand | 13 |
| detached | 12 |
| poultry_house | 9 |
| hotel | 9 |
| shed | 9 |
| garage | 7 |
| office | 7 |
| college | 7 |
| riding_hall | 6 |
| university | 6 |
| boathouse | 6 |
| construction | 5 |
| airport_terminal | 4 |
| hospital | 3 |
| cowshed | 3 |
| fire_station | 2 |
| storage_tank | 2 |
| public | 2 |
| government | 2 |
| pavilion | 1 |
| ruins | 1 |
| sports_hall | 1 |
| prison | 1 |
| service | 1 |
| transportation | 1 |

139 **Supplementary Table S9.** Number of predicted barn polygons that intersect at least one OSM
140 road with the listed tag in the Southeastern states (NC, SC, VA). Multi-tag road entries (e.g.,
141 ['residential', 'service']) indicate polygons intersecting multiple overlapping road features.

| OpenStreetMap© road tag | Count |
|---|---|
| service | 493 |
| motorway | 216 |
| residential | 131 |
| footway | 68 |
| motorway_link | 50 |
| trunk | 25 |
| track | 25 |
| tertiary | 21 |
| secondary | 16 |
| unclassified | 13 |
| ['motorway', 'motorway_link'] | 12 |
| ['motorway_link', 'motorway'] | 12 |
| primary | 10 |
| construction | 8 |
| path | 6 |
| ['residential', 'service'] | 4 |
| ['service', 'residential'] | 4 |
| ['steps', 'footway'] | 3 |
| ['footway', 'steps'] | 3 |
| ['service', 'track'] | 2 |
| ['track', 'service'] | 2 |
| proposed | 1 |
| cycleway | 1 |
| ['service', 'footway'] | 1 |
| ['residential', 'construction'] | 1 |
| steps | 1 |
| bridleway | 1 |
| ['residential', 'unclassified'] | 1 |
| ['unclassified', 'residential'] | 1 |
| trunk_link | 1 |
| ['motorway_link', 'service'] | 1 |
| ['service', 'motorway_link'] | 1 |

142

143  **Supplementary Table S10.** Number of predicted barn polygons that intersect at least one OSM
144  building with the listed tag in the Midwestern states (IA, MN, OH).

| OpenStreetMap© building tag | Count |
|---|---|
| yes | 20511 |
| industrial | 1467 |
| school | 1135 |
| commercial | 725 |
| warehouse | 680 |
| hangar | 425 |
| retail | 396 |
| apartments | 329 |
| barn | 308 |
| farm_auxiliary | 253 |
| garage | 200 |
| shed | 194 |
| university | 136 |
| church | 112 |
| roof | 84 |
| garages | 83 |
| office | 72 |
| greenhouse | 70 |
| house | 57 |
| residential | 55 |
| sty | 48 |
| grandstand | 47 |
| no | 40 |
| college | 40 |
| stable | 35 |
| civic | 24 |
| hotel | 24 |
| farm | 24 |
| prison | 23 |
| dormitory | 22 |
| data_center | 21 |
| public | 17 |
| service | 17 |
| parking | 17 |
| stadium | 16 |
| hospital | 16 |
| manufacture | 16 |
| boathouse | 13 |
| storage | 12 |
| government | 9 |
| detached | 9 |
| transportation | 8 |
| terrace | 6 |
| military | 5 |
| kindergarten | 5 |
| carport | 4 |
| construction | 4 |
| riding_hall | 3 |
| sports_centre | 3 |
| fire_station | 3 |
| agricultural | 3 |
| cowshed | 3 |
| silo | 2 |
| bridge | 2 |
| semidetached_house | 2 |
| yes;commercial | 2 |
| store | 1 |
| commercial;yes | 1 |
| ruins | 1 |
| supermarket | 1 |
| airport_terminal | 1 |
| quonset_hut | 1 |
| train_station | 1 |
| storage_tank | 1 |
| skating_rink | 1 |
| trader | 1 |
| pavilion | 1 |
| roof;yes;warehouse | 1 |
| industrial;yes;roof | 1 |
| chapel | 1 |
| school;roof | 1 |
| allotment_house | 1 |
| motel | 1 |
| post_office | 1 |

145  **Supplementary Table S11.** Number of predicted barn polygons that intersect at least one OSM

146  road with the listed tag in the Midwestern states (IA, MN, OH). Multi-tag road entries (e.g.,

147  ['residential', 'service']) indicate polygons intersecting multiple overlapping road features.

| OpenStreetMap© road tag | Count |
| --- | --- |
| service | 4904 |
| footway | 809 |
| residential | 591 |
| motorway | 477 |
| primary | 308 |
| trunk | 278 |
| tertiary | 264 |
| secondary | 215 |
| motorway_link | 157 |
| unclassified | 122 |
| cycleway | 68 |
| track | 49 |
| pedestrian | 48 |
| path | 45 |
| corridor | 28 |
| steps | 16 |
| trunk_link | 16 |
| primary_link | 9 |
| secondary_link | 8 |
| raceway | 8 |
| proposed | 6 |
| construction | 5 |
| tertiary_link | 4 |
| bridleway | 4 |
| living_street | 2 |
| abandoned | 2 |
| planned | 1 |
| services | 1 |
| road | 1 |
| service | 4904 |
| footway | 809 |
| residential | 591 |

148

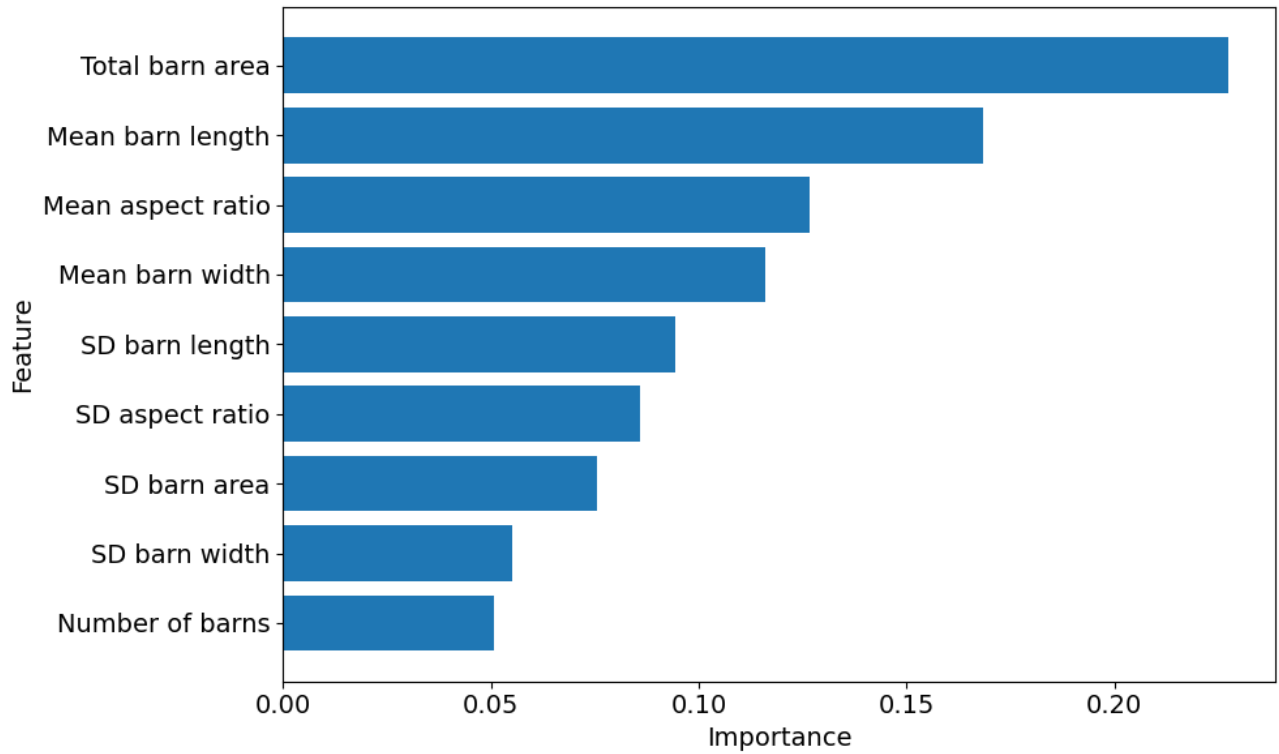

149
150  **Supplementary Material Figure S12. Random Forest feature importance scores for farm-**
151  **level predictors of swine production type.**